\documentclass[11pt,a4paper]{article}
\pdfoutput=1
\usepackage{jheppub}
\usepackage{graphicx}
\usepackage{physics}
\usepackage{listings}
\usepackage{xcolor}
\usepackage[normalem]{ulem}
\usepackage{amsmath}
\usepackage{amsthm}
\usepackage{amssymb}
\usepackage{amsfonts}
\usepackage{enumerate}
\usepackage[utf8]{inputenc}
\usepackage{hyperref}

\usepackage{bbm}
\usepackage{tensor}
\newcommand*{\rep}[2][]{\ensuremath{{\boldsymbol{#2}#1}}}
\newcommand{\birdtrack}[2]{\parbox{#1}{\includegraphics[width=#1]{#2}}}
\newcommand{\stars}[1]{\protect\includegraphics[scale=0.1]{stars/#1.png}}

\definecolor{darkgreen}{HTML}{109930}

\title{The Basis Invariant Flavor Puzzle}
\author[a]{Miguel P. Bento,}
\author[a]{Jo\~ao P. Silva,}
\author[b]{Andreas Trautner}
\affiliation[a]{CFTP, Departamento de F\'{i}sica, Instituto Superior T\'{e}cnico,
Universidade de Lisboa,\\
Avenida Rovisco Pais 1, 1049 Lisboa, Portugal}
\affiliation[b]{
Max-Planck-Institut f\"ur Kernphysik, Saupfercheckweg 1,
69117 Heidelberg, Germany \\
}

\emailAdd{miguel.pedra.bento@tecnico.ulisboa.pt}
\emailAdd{jpsilva@cftp.ist.utl.pt}
\emailAdd{trautner@mpi-hd.mpg.de}

\abstract{
The flavor puzzle of the Standard Model quark sector is formulated in a non-perturbative way, using basis invariants that are independent of the choice of quark field basis.
To achieve this, we first derive the algebraic ring of $10$ CP even (primary) and $1$ CP odd (secondary) basis invariants, using the Hilbert series and plethystic logarithm. An orthogonal basis in the ring of basis invariants is explicitly constructed, using hermitian projection operators derived via birdtrack diagrams.
The thereby constructed invariants have well defined CP transformation behavior and give the most direct access to the flavor symmetric alignments of basis covariants.
We firstly ``measure'' the orthogonal basis invariants from experimental data and characterize their location in the available parameter space.
The experimentally observed orthogonal basis invariants take very close to maximal values and are highly correlated. Explaining the location of the invariants at close to maximal points, including the associated miniscule and highly correlated deviations, corresponds to solving the flavor puzzle in the invariant language. Once properly normalized, the orthogonal basis invariants are close to scale (RGE) invariant, hence, provide exquisite targets for fits of both, low- and high-scale (bottom-up and top-down) flavor models. 
Our result provides an entirely new angle on the flavor puzzle, and opens up ample opportunities for its ultimate exploration.}

\begin{document}
\maketitle

\section{Introduction}
Ever since the theoretical completion of the Standard Model (SM), and especially after the experimental confirmation of all of its constituents, the flavor puzzle 
remains its most captivating mystery~\cite{Weinberg:2020zba}. The question is why there are exactly three generations of matter fermions and what determines 
their pattern of hierarchical masses and intergenerational interactions that also hosts the only definitively observed source of charge-parity (CP) violation in Nature.

The theoretical formulation of the SM flavor sector shows ambiguities in the Higgs Yukawa couplings to the three generations of left- and right-handed 
up and down-type fermions, corresponding to an unphysical choice of basis in the three-generational flavor space $\mathrm{SU(3)}^5$. Since experimental outcomes cannot be affected by 
arbitrary choices of basis or parametrization, physical observables must be given by basis invariant functions that are independent of unphysical basis choices. 
Nonetheless, there exists presently no quantitative investigation of the flavor puzzle exclusively in terms of basis invariant quantities.
The purpose of our paper is to deliver such a formulation of the flavor puzzle entirely in terms of basis invariants, and firstly explore the lessons we can learn from it. 

The current absence of an entirely basis invariant quantitative formulation of the flavor puzzle may be tracked back to technical challenges, 
regarding the questions of \textit{how} a set of minimal basis invariants should be practically and systematically constructed, and 
\textit{when} such a construction is complete. Most well known is certainly the pioneering basis invariant characterization of CP violation in terms
of the so-called Jarlskog invariant~\cite{Greenberg:1985mr,Jarlskog:1985ht}, which also has been extended to enlarged fermion and scalar sectors~\cite{Bernabeu:1986fc,Branco:1986gr,Botella:1994cs,Lavoura:1994fv,Yu:2020gre}. 
But also the complete ring of  basis invariants of the SM (and most common neutrino sector extensions) 
has been constructed, using the Hilbert series of invariant theory~\cite{Jenkins:2007ip,Jenkins:2009dy,Hanany:2010vu}. 

Since any algebraic combination of basis invariants is itself a basis invariant, one might think that there are 
no meaningfully discernible bases in the space of invariants. Perhaps because of this, little attention has been paid 
to the explicit and systematic construction of the basis invariants themselves. %, which typically leads to a more complicated treatment than necessary. 
However, using an arbitrary basis in the space of invariants can lead to more complicated expressions than necessary.
Hence, we pay close attention to the systematic construction of the basis invariants themselves.
We will follow the strategy outlined in~\cite{Trautner:2018ipq,Trautner:2020qqo}, using hermitian projection operators~\cite{Keppeler:2013yla} (see also~\cite{Alcock-Zeilinger:2016cva,Alcock-Zeilinger:2016sxc,Alcock-Zeilinger:2016xgf}), in order to systematically construct an orthogonal basis in the ring of basis invariants.\footnote{%
We stress that an orthogonal basis for an algebraic ring is quite different from an orthogonal basis of a linear vector space.
For example, an orthogonal basis for a ring cannot simply be infinitesimally rotated to obtain another orthogonal basis.
In the most general sense, our notion of orthogonality is based on the mutually vanishing products of different projection operators that project covariantly transforming irreducible representations (incl.\ trivial singlets) out of tensors of, in principle, arbitrary rank.}

The necessary projection operators can conveniently be constructed using birdtrack diagrams~\cite{Cvitanovic:1976am,Cvitanovic:2008zz,Keppeler:2017kwt}.
The thereby derived projection operators allow to project arbitrary rank tensors onto their irreducible covariant contents, incl.\ trivial singlets i.e.\ invariants, which allows us to track down the origin of a given invariant to independent covariant channels. 
The thus obtained orthogonal invariants and their relations are as short as possible by construction. As an additional benefit, the formulation in terms of basis invariants simplifies the analysis of renormalization group equations (RGE), RGE running and the derivation of RGE invariants~\cite{Harrison:2010mt}, which has been observed for both, SM fermion~\cite{Feldmann:2015nia, Chiu:2015ega, Chiu:2016qra, Yu:2020gre, Wang:2021wdq}, as well as for extended scalar sectors~\cite{Herren:2017uxn, Bednyakov:2018cmx, Bijnens:2018rqw} and is expected to be even further simplified if the involved invariants are orthogonal to each other.

Moreover, using basis invariants is a powerful tool to examine the violation of CP and other global symmetries, see~\cite{Ivanov:2018ime, Ivanov:2019kyh,deMedeirosVarzielas:2019rrp,Bento:2020jei}.
Symmetries manifest themselves either in the vanishing of non-trivially transforming covariants, or in alignment of covariants that correspond to specific relation of basis invariants~\cite{Bento:2020jei}.
Using orthogonal invariants, such relations become as short and transparent as possible, and this may also help to detect symmetries and their violation in the SM.

Our principal technique to construct orthogonal invariants generally applies to both, quark and lepton sectors. 
However, the different possible mechanisms of neutrino mass generation involve different covariant tensor structures, hence, different possibilities for invariant rings of the lepton sector, see~\cite{Hanany:2010vu,Yu:2021cco,Wang:2021wdq}. 
For this reason, we entirely focus on the quark sector here which is free of such ambiguities, and whose parameters have been experimentally determined with very high precision.
The main task is to derive an orthogonal basis of flavor invariants, after which we quantitatively examine them in order to obtain the complete basis and parametrization invariant picture of the SM flavor puzzle.

The paper is structured as follows. Section~\ref{sec:invariants} gives a general overview over the SM quark sector flavor covariants and different parametrizations used to evaluate them. 
We also state our choice of orthogonal basis invariants, as well as the syzygy that relates our ten primary invariants to the CP-odd, secondary invariant known as the Jarlskog invariant. Subsequently, in~\ref{sec:hilbert_series} we formally characterize the invariant ring of the SM quark sector using the Hilbert series and plethystic logarithm. This is followed by the construction of our orthogonal, adjoint space basis of projection operators in sec.~\ref{sec:projector_basis}, which have been used to construct the orthogonal invariants of sec.~\ref{sec:invariants}. The CP transformation behavior of basis co- and invariants is unveiled in section~\ref{sec:CP_trafo}. Finally, we quantitatively analyze the parameter space of the orthogonal invariants, determine their experimental values and errors, as well as their renormalization group evolution in sections~\ref{sec:quantitative} and~\ref{sec:renormalization}, respectively. Sections~\ref{sec:discussion} and~\ref{sec:conclusions} contain further discussions and comments, as well as our conclusions. In six appendices, we provide 
\ref{app:CP-even_ring} a discussion of the CP-even subring of the SM, \ref{sec:birdtracks} useful birdtrack identities, \ref{sec:normalization} a comment about the group theoretically correct and unitary normalization factors of the projection operators, \ref{App:Ialt} plots for an alternative normalization of the invariants, \ref{app:inner_product} the Frobenius inner product to set limits on the boundaries of the invariant parameter space, and \ref{app:CKMrunning}, an up-to-date display of the running CKM parameters. 

\section{Quark sector flavor invariants}\label{sec:invariants}
The Yukawa couplings of the SM quark sector are given by the Lagrangian
\begin{equation}\label{eq:SMYukawas}
 -\mathcal L_\mathrm{Yuk.}~=~\bar{Q}_{\mathrm{L},i}\,\widetilde{H}\,{\left[Y_u\right]}\indices{^i_{j}}\,u^{j}_\mathrm{R}+\bar{Q}_{\mathrm{L},i}\,H\,{\left[Y_d\right]}\indices{^i_{k}}\,d_\mathrm{R}^{k}
%+\bar{L}^{i}H\,y^{ij}_e\,e_\mathrm{R}^{j}+\bar{L}^{i}\widetilde{H}\,y^{ij}_\nu\,\nu_\mathrm{R}^{j}
 +\mathrm{h.c.}\;,
\end{equation}
where we explicitly display the flavor indices $i,j,k=1,2,3$. The Yukawa coupling matrices $Y_u$ and $Y_d$ are general complex $3 \times 3$ matrices. 
Under general flavor space redefinitions of the quark fields, described by the group $\mathrm{SU}(3)_{Q_\mathrm{L}}\otimes\mathrm{SU}(3)_{u_\mathrm{R}}\otimes\mathrm{SU}(3)_{d_\mathrm{R}}$, the Yukawa matrices transform covariantly as \mbox{$Y_u\mathrel{\hat=}(\rep{\bar{3}},\rep{3},\rep{1})$} and $Y_d\mathrel{\hat=}(\rep{\bar{3}},\rep{1},\rep{3})$. Hence, $Y_u$ and $Y_d$ can be regarded as \textit{spurions} of 
flavor symmetry breaking. We define the two matrices
\begin{equation}
    \widetilde H_u := Y_u Y_u^\dagger \, , \quad
    \widetilde H_d := Y_d Y_d^\dagger \, .
\end{equation}
$\widetilde{H}_{u,d}$ are hermitian with positive eigenvalues. Taking these products, the right-handed spaces are traced out\footnote{%
Since there is exactly one triplet covariant for each of the right-handed quark flavor spaces, there is only exactly one invariant that 
can be constructed for each of the right-handed spaces. These two invariants are given by $\Tr(Y_u^\dagger Y_u)$ and $\Tr(Y_d^\dagger Y_d)$, 
and since they are equivalent to $\Tr \widetilde H_u$ and $\Tr \widetilde H_d$ they are automatically included in our treatment.}
and $\widetilde H_u$ and $\widetilde H_d$ each transform as $\rep{\bar{3}}\otimes\rep{3}=\rep{1}\oplus\rep{8}$
in the left-handed quark flavor space.
The singlet pieces are given by $\Tr \widetilde H_u$ and $\Tr \widetilde H_d$. Hence, the octet pieces can be isolated as~\cite{Fonseca}
\begin{equation}\label{eq:traceless}
 H_u:= \widetilde H_u - \mathbbm{1} \Tr \frac{\widetilde H_u}{3} \qquad\text{and}\qquad
H_d:= \widetilde H_d - \mathbbm{1} \Tr \frac{ \widetilde H_d}{3}\;.
 \end{equation}
$H_u$ and $H_d$ are traceless hermitian matrices with eight parameters each. 

There are $10$ physical parameters in the quark Yukawa sector ($3$ up-type masses, $3$ down-type masses, $1$ CP odd and $3$ CP even mixing parameters~\cite{Santamaria:1993ah}). 
The physical basis corresponds to diagonal kinetic terms after spontaneous symmetry breaking, and is achieved by bi-unitarily diagonalizing the Yukawa matrices 
as 
\begin{align}
 &V_{u,\mathrm{L}}^\dagger \,Y_u\, V_{u,\mathrm{R}}~=~\frac{\sqrt{2}}{v}\;\mathrm{diag}(m_u, m_c, m_t)\qquad\text{and} \\
 &V_{d,\mathrm{L}}^\dagger \,Y_d\, V_{d,\mathrm{R}}~=~\frac{\sqrt{2}}{v}\;\mathrm{diag}(m_d, m_s, m_b)\;.
\end{align}
For clarity, we express this in terms of the masses here, such that dividing out the Higgs vacuum expectation value $v=246\,\mathrm{GeV}$
yields the dimensionless diagonal Yukawa couplings. In the physical basis, the covariant tensors can be expressed as
\begin{align}\label{eq:HuPhysical}
 &\widetilde H_u~=~\mathrm{diag}(\,y_u^2\,,\,y_c^2\,,\,y_t^2\,)\qquad\qquad \\\label{eq:HdPhysical}
 \text{and} \qquad &\widetilde H_d~=~V_{\mathrm{CKM}}\;\mathrm{diag}(\,y_d^2\,,\, y_s^2\,,\, y_b^2\,)\;V_{\mathrm{CKM}}^\dagger\;,
\end{align}
where $V_{\mathrm{CKM}}:=V_{u,\mathrm{L}}^\dagger V_{d,\mathrm{L}}$ is the Cabibbo-Kobayashi-Maskawa (CKM) matrix~\cite{Kobayashi:1973fv}.
In the standard parameterization it can be written as~\cite{ParticleDataGroup:2022pth} 
\begin{equation}
    V_{\mathrm{CKM}} =
    \begin{pmatrix}
    1 & 0 & 0 \\
    0 & c_{23} & s_{23} \\
    0 & - s_{23} & c_{23}
    \end{pmatrix}
    \begin{pmatrix}
    c_{13} & 0 & s_{13} e^{-\mathrm{i} \delta} \\
    0 & 1 & 0 \\
    - s_{13} e^{\mathrm{i} \delta} & 0 & c_{13}
    \end{pmatrix}
    \begin{pmatrix}
    c_{12} & s_{12} & 0 \\
    - s_{12} & c_{12} & 0 \\
    0 & 0 & 1
    \end{pmatrix} \, ,
\end{equation}
where the quark mixing angles appear as $s_{ij}=\sin\theta_{ij}$ and $c_{ij}=\cos\theta_{ij}$, and CP violation is parametrized by the complex phase $\delta$.
For concrete applications, it is often convenient to use the Wolfenstein parametrization of the CKM matrix~\cite{Wolfenstein:1983yz}. Since it is better suited for our purpose, we adopt the Wolfenstein-like but exactly unitary parametrization of~\cite{Buras:1994ec}, for which we obtain up-to-date values for the parameters $\lambda=0.22481\pm0.00059$, $A=0.817\pm0.018$, $\rho=0.145\pm0.015$ and $\eta=0.366\pm0.012$
by performing our own fit to PDG data~\cite{ParticleDataGroup:2022pth}.

The $10$ physical parameters correspond to $10$ algebraically independent\footnote{%
Algebraic \mbox{(in-)dependence} of invariants is easily tested by numerically evaluating the rank of their Jacobi matrix, see e.g.~\cite[Appendix~A]{Trautner:2018ipq}.} 
\textit{primary} basis invariants that must be given as functions of $\widetilde H_u$ and $\widetilde H_d$.
Various choices for the ten primary invariants are possible. We will first state our set of primary basis invariants, then subsequently motivate our choice
and outline the systematic construction of our basis invariants in the following sections.

We have already identified the two ``trivial'' primary invariants 
\begin{equation}\label{eq:trivialInvariants}
I_{10}:=\Tr \widetilde H_u\qquad\text{and}\qquad I_{01}:=\Tr \widetilde H_d\;.
\end{equation}
Hence, eight more algebraically independent basis invariants have to arise as trivial singlets in the tensor products $\rep{8}^{\otimes n}_u\otimes\rep{8}^{\otimes m}_d$ of 
the $\rep{8}$-plet covariant tensors $H_u$ and $H_d$. We construct those eight algebraically independent primary basis invariants as
\begin{align}\notag
    &I_{20} := \Tr (H_u^2)\,,\quad  I_{02} := \Tr (H_d^2)\,,\quad I_{11} := \Tr (H_u H_d)\,,&  \\[0.2cm] \label{eq:trace_basis_inv}
    &I_{30} := \Tr (H_u^3) \,,\quad I_{03} := \Tr (H_d^3)\,,\quad I_{21} := \Tr (H_u^2 H_d)\,,\quad I_{12} := \Tr (H_u H_d^2) \, ,& \\[0.2cm] \notag
    &I_{22} := 3 \Tr (H_u^2 H_d^2) - \Tr (H_u^2)\Tr (H_d^2)\,.&
\end{align} 
The above $10$ basis invariants correspond to the $10$ physical parameters of the SM quark Yukawa sector.
It is intuitively clear that the up and down sector masses must correspond to $I_{10}$, $I_{20}$, $I_{30}$, and 
$I_{01}$, $I_{02}$, $I_{03}$, respectively. Furthermore, the \mbox{(mis-)alignment} of up and down sectors, i.e.\ CKM angles and phase, 
must correspond to the mixed invariants $I_{11}$, $I_{21}$, $I_{12}$ and $I_{22}$, in agreement with~\cite{Branco:1987mj}.
Explicit expressions for the masses and CKM squared elements can be obtained as a combination of invariants, see e.g.~\cite{Talbert:2021iqn,Bree:2023ojl}, but this is not 
the topic of the present paper where we set out to characterize the orthogonal invariants themselves.

All of the $10$ primary basis invariants are real and CP even, as we will derive in detail in section~\ref{sec:CP_trafo}. Hence, one may wonder how CP violation is encoded in them.
It is well known that the only CP-odd basis invariant that can be constructed in the SM (ex.~$\bar\Theta$) is given by the so-called Jarlskog invariant~\cite{Greenberg:1985mr,Jarlskog:1985ht}.
The Jarlskog invariant can be defined as
\begin{equation}\label{eq:Jdefinition}
 J_{33} := \Tr (H_u^2 H_d^2 H_u H_d) - \Tr (H_d^2 H_u^2 H_d H_u) \equiv \frac{1}{3} \Tr \left[ H_u, H_d \right]^3 \,.
\end{equation}
In the physical basis, and using the standard parametrization, it is given by
\begin{equation}\label{eq:J33}
    J_{33} = \mathrm{i}\,J\,\frac{2^7}{v^{12}}\,\left(m_t^2 - m_c^2\right)\,\left(m_t^2 - m_u^2\right)\,\left(m_c^2 - m_u^2\right)\,\left(m_b^2 - m_s^2\right)\,\left(m_b^2 - m_d^2\right)\,\left(m_s^2 - m_d^2\right)\,,
\end{equation}
where 
\begin{equation}\label{eq:J}
    J := \cos\theta_{12} \, \cos^2\theta_{13}\, \cos\theta_{23}\,\sin\theta_{12}\,\sin\theta_{13}\,\sin\theta_{23}\,\sin\delta\approx A\,\eta\,\lambda^6\,.
\end{equation}
The Jarlskog invariant is not included in the set of primary invariants above, but arises as a \textit{secondary} invariant in the ring of basis invariants. This implies that it is not
algebraically independent of the CP-even primary invariants but fulfills a polynomial relation with them.
As we will see below, this relation is the syzygy of the ring of invariants. It is given by 
\begin{align}
    \left(J_{33}\right)^2 =& -\frac{4}{27} I_{22}^3 + \frac{1}{9} I_{22}^2 I_{11}^2 + \frac{1}{9} I_{22}^2 I_{02} I_{20} 
    +\frac{2}{3}I_{22}I_{30}I_{03}I_{11} -\frac{2}{3}I_{22}I_{21}I_{12}I_{11}-\frac{1}{9}I_{22}I_{11}^2I_{20}I_{02}  \nonumber \\[2mm]
    &+\frac{2}{3}I_{22}I_{21}^2I_{02}+\frac{2}{3}I_{22}I_{12}^2I_{20}-\frac{2}{3}I_{22}I_{30}I_{12}I_{02}-\frac{2}{3}I_{22}I_{03}I_{21}I_{20}  \nonumber \\[2mm]
    &-\frac{1}{3}I_{30}^2I_{03}^2+I_{21}^2I_{12}^2+2I_{30}I_{03}I_{21}I_{12}-\frac{4}{9}I_{30}I_{03}I_{11}^3 \nonumber \\[2mm]
    &+\frac{1}{18}I_{30}^2I_{02}^3+\frac{1}{18}I_{03}^2I_{20}^3-\frac{4}{3}I_{30}I_{12}^2-\frac{4}{3}I_{03}I_{21}^2  \nonumber \\[2mm]
    &-\frac{1}{3}I_{30}I_{21}I_{11}I_{02}^2-\frac{1}{3}I_{03}I_{12}I_{11}I_{20}^2+\frac{2}{3}I_{30}I_{12}I_{11}^2I_{02}+\frac{2}{3}I_{03}I_{21}I_{11}^2I_{20}\nonumber \\[2mm] \label{eq:syzygy}
    &-\frac{2}{3}I_{21}I_{12}I_{20}I_{02}I_{11}-\frac{1}{108}I_{20}^3I_{02}^3 +\frac{1}{36}I_{20}^2I_{02}^2I_{11}^2
    +\frac{1}{6}I_{21}^2I_{20}I_{02}^2+\frac{1}{6}I_{12}^2I_{02}I_{20}^2\;.
\end{align}
For our choice of primary invariants, this relation is described by a polynomial of 27 terms (out of 37 possible power products of lower lying non-trivial invariants, not involving the trivial invariants $I_{10}$ and $I_{01}$).
This should be compared to the result of Jenkins and Manohar~\cite{Jenkins:2009dy} which involved $241$ terms in order to express $J_{33}^2$ in terms of their choice of CP-even invariants. The invariants of~\cite{Jenkins:2009dy} are very similar to our choice in~\eqref{eq:trace_basis_inv} but involve the traceful $\widetilde{H}_u$ and $\widetilde{H}_d$.
The enormous simplification of the syzygy, hence, arises from our usage of \textit{orthogonal} basis invariants, constructed from \textit{orthogonal} projection operators, as we will further elaborate on in section~\ref{sec:projector_basis}. The usage of orthogonal projection operators accomplishes to isolate irreducible representations out of arbitrary (potentially highly non-linear) tensor structures. For example, the orthogonality of invariants in the adjoint space of the SM quark flavor ring 
\textit{automatically} ensures the removal of the traces of $\widetilde{H}_u$ and $\widetilde{H}_d$ (singlet pieces in $\rep{1}\oplus\rep{8}$)  in all non-linear invariants, as had to be performed by hand in eq.~\eqref{eq:traceless}.
Also the specific choice of $I_{22}$ \eqref{eq:trace_basis_inv} is motivated by orthogonality and the fact that amongst all possible orthogonal quartic invariants (of which we will see there are multiple possibilities) 
our choice gives rise to the shortest syzygy.

\section{Construction of an orthonormal basis of flavor invariants}\label{sec:construction}
We will now formalize the construction of our set of orthonormal primary and secondary invariants
stated in the previous section. For this, we first construct the structure of the ring of basis invariants
using the Hilbert series and plethystic logarithm. Subsequently, we explicitly construct the invariants using
orthogonal hermitian projection operators that are constructed using the technique of birdtrack diagrams.

\subsection{The Hilbert series}\label{sec:hilbert_series}
The Hilbert series (HS) of the SM quark sector is straightforwardly calculated (see~\cite{Lehman:2015via} for a concise introduction to the HS technique). 
The covariantly transforming objects $H_u$ and $H_d$ are $\rep{8}$-plets under the left-handed 
$\mathrm{SU}(3)$ flavor rotation. Hence, the HS is computed via the integral
\begin{equation}\label{eq:hilbert_integral}
    H(K[V]^G; u, d) = \int_{\mathrm{SU}(3)} d \mu_{\mathrm{SU}(3)} \,
    \mathrm{PE}\left[ z_1, z_2 ; u; \mathbf{8} \right]
    \mathrm{PE}\left[ z_1, z_2 ; d; \mathbf{8} \right] \, ,
\end{equation}
where the plethystic exponential $\mathrm{PE}\left[ z_1, z_2 ; u; \mathbf{8} \right]$ and the integral measure can be found in~\cite[Eq.~5.6]{Bento:2021hyo} (see also~\cite{Benvenuti:2006qr,Feng:2007ur,Gray:2008yu, Hanany:2008sb, Lehman:2015via}).

We are working in a parameter space with dimension $\dim V = 16$, transforming under $G = \mathrm{SU}(3)$ with dimension $\dim G = 8$.
Hence, the number of physical parameters is given by
\begin{equation}
    N_\text{physical} = \dim V - \dim G = 8 \, .
\end{equation}
Together with the two trivial invariants this reproduces the number of $10$ physical parameters.
The integral in eq.~\eqref{eq:hilbert_integral} is straightforwardly computed to yield the multi-graded HS
\begin{equation}\label{eq:hilbert_graded}
    H(K[V]^G; u, d) = \frac{1+u^3 d^3}
    {(1 - u^2) (1 - d^2) (1 - ud) (1 - u^3) (1 - d^3)
    (1 - ud^2) (1 - u^2 d) (1 - u^2 d^2)}\,.
\end{equation}
The ungraded form of the HS (setting the dummy indices as $t=u=d$) is directly read off as
\begin{equation}\label{eq:hilbert_ungraded}
    H(K[V]^G, t) = \frac{1+t^6}{(1 - t^2)^3 (1 - t^3)^4 (1 - t^4)} \, .
\end{equation}
The denominators in eqs.~\eqref{eq:hilbert_graded}--\eqref{eq:hilbert_ungraded}
determine the number and order of primary invariants. The numerators give the secondary invariants. 
The multi-graded HS additionally informs about the structure of the invariant in terms of the covariants. 

Another function of interest is the plethystic logarithm (PL)~\cite{getzler_kapranov_1998,labastida} (in the particle physics context it was first introduced and used in~\cite{Benvenuti:2006qr,Feng:2007ur,Butti:2007jv,Gray:2008yu,Hanany:2008sb}), defined as 
\begin{equation}
 \mathrm{PL} \left[ H(K[V]^G; u, d) \right]:=\sum\limits_{k=1}^{\infty}\frac{\mu(k)\, \ln H(K[V]^G; u^k, d^k)}{k}\;,
\end{equation}
where $\mu(k)$ is the M\"obius function. The PL of our ring can be computed exactly, because it terminates at order $u^6d^6$. It is given by 
\begin{equation}\label{eq:plethystic_log}
    \mathrm{PL} \left[ H(K[V]^G; s, t) \right] =
   u^2 + u d + d^2 + u^3 + d^3 + u^2 d + u d^2 + u^2 d^2 + u^3 d^3 - u^6 d^6 \,.
\end{equation}
The leading positive terms of the PL correspond to the number and structure of the generating set of (primary and secondary) invariants of the ring.
The final negative term cuts off the generating set and informs us that there is a syzygy of order $u^6d^6$ between the invariants of the generating set.
For our choice of primary and secondary invariants we have already explicitly stated this syzygy in eq.~\eqref{eq:syzygy}.\footnote{For the construction
of general invariant relations and syzygies we refer to the procedure outlined in~\cite{Trautner:2018ipq} (that was adopted also in~\cite[App.~C]{Wang:2021wdq}).}

Our choice of primary and secondary invariants conveniently realize the ring in form of a Hironaka decomposition \cite{Hochster:1974a, Hochster:1974b} (see e.g.\ also~\cite[Sec.~2.3]{Sturmfels:2008}, \cite[Sec.~5.4.1]{Henning:2017fpj}). Therefore, it is guaranteed that \textit{any} basis invariant quantity (any observable) $\mathcal{O}_\mathrm{flavor}$ of the SM quark sector can be expressed through our generating set of invariants as
\begin{equation}\label{eq:hironaka_flavor}
    \mathcal{O}_\mathrm{flavor} = \mathbbm{C}[I] + J_{33}\,\mathbbm{C}[I]\,,
\end{equation}
where $\mathbbm{C}[I]$ denote polynomials in the primary invariants $I$ with potentially complex coefficients.

Lastly, we note that it is possible and instructive to formulate an $\mathrm{SO}(3)$ version of the SM ring in the absence of CP violation, an exercise that we perform in appendix~\ref{app:CP-even_ring}.

\subsection{Construction of orthogonal invariant projection operators}\label{sec:projector_basis}
Given the number and structure of primary and secondary invariants as obtained from the HS and PL, we proceed with the explicit construction of invariants using projection operators.
The fact that any algebraic combination of basis invariants yields another basis invariant 
invites the common misconception that there are no meaningfully distinguishable bases in the space of invariants.
However, if invariants are obtained using projection operators, then properties such as orthogonality 
of invariants can be defined based on the orthogonality of the respective projection operators. 
A good choice of basis should be an orthogonal basis. There can be different possible constructions of orthogonal projection operators, hence, different orthogonal bases, potentially suitable for different applications. 

As a first step in the quantitative basis independent exploration of the SM flavor puzzle, 
we construct here an orthogonal set of SM flavor invariants in the adjoint space of flavor. 
This explicitly shows that the trace basis invariants of section~\ref{sec:invariants} arise as an orthogonal 
basis of invariants in the adjoint space of left-handed quark flavor. 

In left-handed quark flavor space, $\widetilde{H}_u$ and $\widetilde{H}_d$ transform as $\rep{\bar{3}}\otimes\rep{3} = \rep{8}\oplus\rep{1}$.
Graphically, in terms of birdtrack diagrams~\cite{Cvitanovic:1976am, Cvitanovic:2008zz} (for a concise introduction, see~\cite{Keppeler:2017kwt}), this relation is represented via projection operators 
of $\rep{\bar{3}}\otimes\rep{3}\rightarrow\rep{\bar{3}}\otimes\rep{3}$ as
\begin{equation}\label{eq:schouten}
 \birdtrack{16ex}{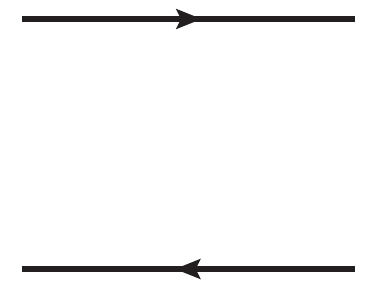}\quad=\quad\frac{1}{N}\birdtrack{16ex}{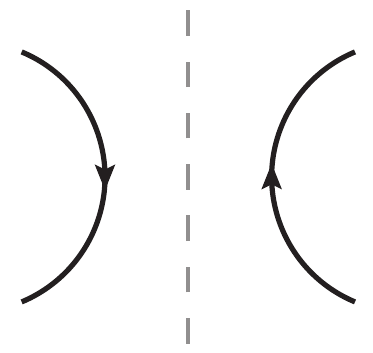}\quad+\quad\frac{1}{T_{\rep{r}}}\birdtrack{16ex}{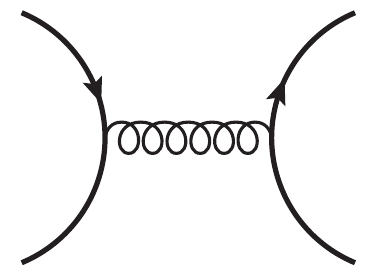}\;.
\end{equation}
Solid lines correspond to contractions in the fundamental space of flavor ($N=3$ for $\mathrm{SU}(N)$), while wavy lines
correspond to contractions in the adjoint space. $T_{\rep{r}}$ for a representation $\rep{r}$ is defined via $\Tr(t^at^b)=T_{\rep{r}} \delta^{ab}$,
and we collect other useful identities for birdtrack computations in appendix~\ref{sec:birdtracks}.
The Lie-algebra generators $(t^{a})^{i}_{\,j}$ ($a,b,\ldots=1,\ldots,8$ and $i,j,\ldots=1,\ldots,3$) of the flavor transformation
are graphically represented as 
\begin{equation}
 (t^a)^i_{\,j}~=~\raisebox{6pt}{\birdtrack{16ex}{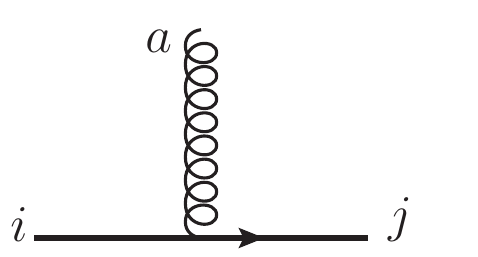}}\;.
\end{equation}
The factorization of the corresponding projection operator in \eqref{eq:schouten} (as emphasized by the dashed line) unambiguously signals the ``nucleation'' of a basis invariant~\cite{Trautner:2018ipq}.
The trivial singlet components of $H_u$ and $H_d$, hence, are given by 
\begin{equation}
 \Tr[\widetilde{H}_u]=\birdtrack{9ex}{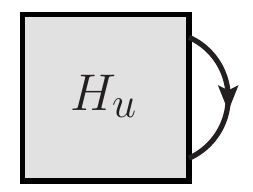}\;\qquad\text{and}\qquad\Tr[\widetilde{H}_d]=\birdtrack{9ex}{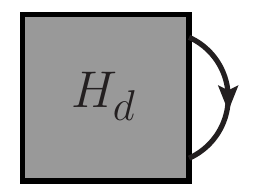}\;.
\end{equation}
Likewise, the adjoint space components of $H_u$ and $H_d$ are obtained via the projections\footnote{%
From here on it makes no difference whether we work with $\widetilde{H}_{u,d}$ or their trace-subtracted counterparts $H_{u,d}$, as the projection operators
\textit{automatically} pick the orthogonal (i.e.\ traceless) components of $\widetilde{H}_{u,d}$.}
\begin{align}
 \boldsymbol{u}^a &\,=\,\Tr[Y_u^\dagger\,t^a\,Y_u]=\birdtrack{16ex}{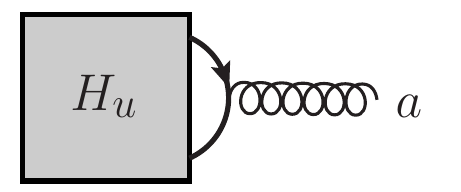}\qquad\text{and}& \\ 
 \boldsymbol{d}^a &\,=\,\Tr[Y_d^\dagger\,t^a\,Y_d]=\birdtrack{16ex}{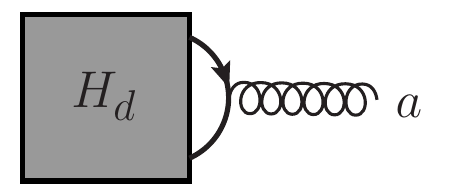}\;.&
\end{align}
These $\rep{8}$-plet vectors in adjoint space are the objects entering the Hilbert series in section~\ref{sec:hilbert_series}.

In order to explicitly construct the resulting invariants, we have to construct orthogonal projection operators in adjoint space
for a mapping of $k$ objects, transforming in adjoint space, onto trivial singlets, 
\begin{equation}
 \rep{8}^{\otimes k}\,\longrightarrow\,\mathbbm{C}\;.
\end{equation}
Performed systematically, this involves the construction 
of all $\rep{8}^{\otimes k}\,\rightarrow\,\rep{8}^{\otimes k}$ projection operators, then selecting the ones
that factorize to yield the trivial singlet irreducible representations. Precisely because of their necessary factorization, 
this procedure can be abridged if only trivial singlet projection operators are sought after.
For example, the operators for $\rep{8}^{\otimes4}\rightarrow\mathbbm{C}$
are readily obtained from \textit{all} projection operators of $\rep{8}^{\otimes2}\rightarrow\rep{8}^{\otimes2}$. This includes operators that project
the direct product of two eights onto the irreducible representations 
\begin{equation}\label{eq:88irreps}
\rep{8}^{\otimes2}=\rep{1}\oplus\rep{8}_\mathrm{S}\oplus\rep{8}_\mathrm{A}\oplus\rep{10}\oplus\overline{\rep{10}}\oplus\rep{27}\;, 
\end{equation}
but also the so-called transition operators~\cite{Keppeler:2012ih,Alcock-Zeilinger:2016cva}, which here correspond to transitions $\rep{8}_\mathrm{S}\leftrightarrow\rep{8}_\mathrm{A}$ between identically transforming irreps in the direct sum on the r.h.s.
Together, these operators form a complete orthogonal basis of all $k$-legged tensor structures, where in above case $k=4$.
Retrieving the singlet projection operators in $\rep{8}^{\otimes4}\rightarrow\rep{8}^{\otimes4}$ from all projection operators in 
$\rep{8}^{\otimes2}\rightarrow\rep{8}^{\otimes2}$ then merely corresponds to formally re-assigning the legs of the respective operator
and re-adjusting its normalization, as we discuss in detail in appendix~\ref{sec:normalization}. Finally, contracting each of the legs with covariantly transforming objects in adjoint space -- here all possible combinations of $\boldsymbol{u}^a$ and $\boldsymbol{d}^a$ -- then yields the individual orthogonal invariants.

The necessary adjoint space projection operators are summarized in the following; below we discuss their normalization. 
For $\rep{8}^{\otimes2}\rightarrow\mathbbm{C}$ the only projection operator is
\begin{equation}\label{eq:dab}
 \delta^{ab}=\birdtrack{16ex}{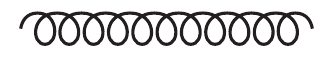}\;.
\end{equation}
For $\rep{8}^{\otimes3}\rightarrow\mathbbm{C}$ there are two orthogonal invariant structures given by 
\begin{equation}\label{eq:cubic}
 \birdtrack{12ex}{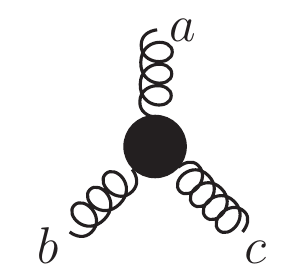}=\mathrm{i}\,f^{abc}\;\qquad\text{and}\qquad\birdtrack{12ex}{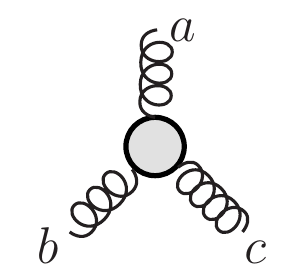}=d^{abc}\;.
\end{equation}
Here,
\begin{align}
& f^{abc}=\frac{1}{\mathrm{i}\,T_{\rep{r}}} \Tr\left(\left[t^a,t^b\right]t^c\right)& &\text{and}&
& d^{abc}=\frac{1}{T_{\rep{r}}} \Tr\left(\left\{t^a,t^b\right\}t^c\right) \;&
\end{align}
are the anti-symmetric and symmetric, respectively, invariant tensors of $\mathrm{SU}(N)$ where 
square (curly) brackets $[\cdot,\cdot]$ ($\{\cdot,\cdot\}$) denote the \mbox{(anti-)commutator}.

For $\rep{8}^{\otimes4}\rightarrow\mathbbm{C}$ there are eight invariant structures given by~\cite{MacFarlane:1968vc,Keppeler:2012ih} (see also \cite[Tab.~9.4]{Cvitanovic:2008zz})
\begin{gather}\label{eq:P4first}
 \rep{1}:\quad\birdtrack{12ex}{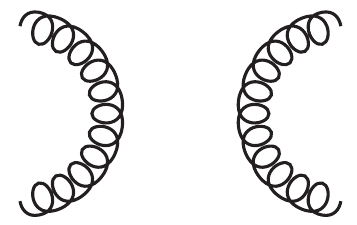}  \\  
\begin{aligned} 
 \rep{8}_\mathrm{S}:& \quad \birdtrack{16ex}{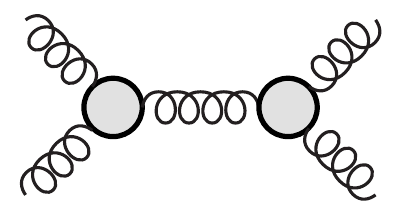}& \qquad\qquad \rep{8}_\mathrm{A}:&\quad\birdtrack{16ex}{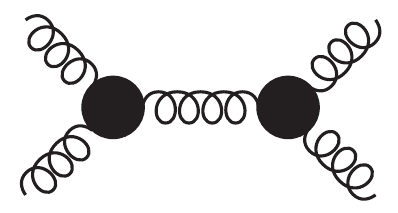}&  \\ \label{eq:8sa}
 \rep{8}_{\mathrm{A}\rightarrow\mathrm{S}}:& \quad \birdtrack{16ex}{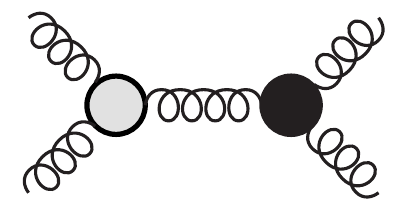}& \qquad \rep{8}_{\mathrm{S}\rightarrow\mathrm{A}}:&\quad \birdtrack{16ex}{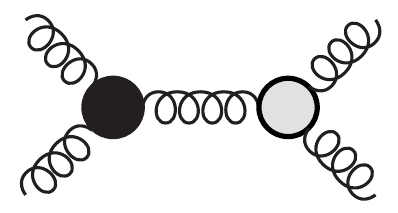}&
\end{aligned} 
\\\label{eq:P4last}
 \birdtrack{16ex}{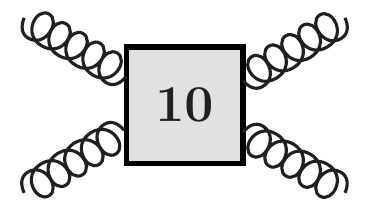} \qquad \birdtrack{16ex}{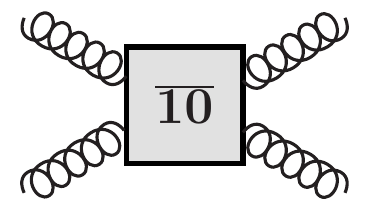} \qquad \birdtrack{16ex}{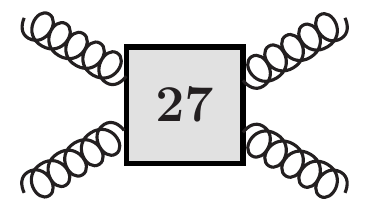}\;.
\end{gather}
Viewed as maps in $\rep{8}^{\otimes2}\rightarrow\rep{8}^{\otimes2}$, these are the projection operators realizing the decomposition into irreps in~\eqref{eq:88irreps},
as well as the $\rep{8}_\mathrm{S}\leftrightarrow\rep{8}_\mathrm{A}$ transition operators in~\eqref{eq:8sa}.
The explicit internal structure of the $\rep{10}$, $\overline{\rep{10}}$, and $\rep{27}$ projection operators can be found in~\cite[Eq~1.23]{Keppeler:2012ih}.

For $\rep{8}^{\otimes5}\rightarrow\mathbbm{C}$ there are no new non-trivial (in the sense of non-factorizing) invariant structures in the SM, as witnessed also by the absence of the respective term in the HS. 

For $\rep{8}^{\otimes6}\rightarrow\mathbbm{C}$ several orthogonal operators exist but we do not display the complete basis of projection operators here as they are not all necessary. The non-trivial, orthogonal projection operator
\begin{equation}\label{eq:Jarlskog}
\birdtrack{24ex}{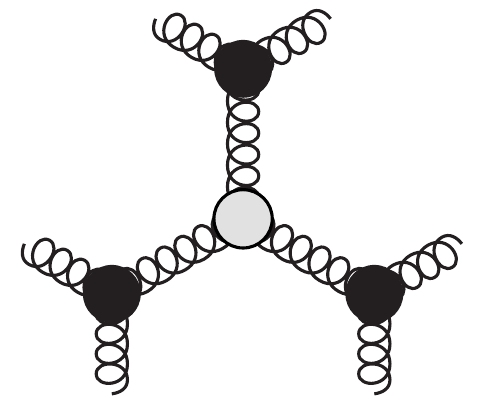}
\end{equation}
can be used to construct the Jarlskog invariant, which is the only new invariant at this level. This completes the set of necessary projection operator structures.

All of the above operators are orthogonal to each other, hence, produce orthogonal basis invariants when contracted with $\boldsymbol{u}^a$'s and/or $\boldsymbol{d}^a$'s.
This is straightforward to show using the diagrammatic identities collected in appendix~\ref{sec:birdtracks}.
We have displayed the operators at this stage without normalization factors.
Different normalizations can make sense depending on how the respective operator is interpreted.
Knowing the correct normalization is not necessary to 
extract the ``essence'' of the corresponding invariant (i.e.\ the invariant with correct relative prefactors of all terms but with an arbitrary global prefactor).
However, the correct normalization of the operators must be known in order to obtain the correct absolute values of the invariants, 
including all group theoretical global prefactors.

In general, projection operators ought to be idempotent ($P^2=P$) and this fixes the normalization, 
with $\Tr P$ being equal to the dimension of the space the operator projects onto. 
For the projection operators for $\rep{8}^{\otimes4}\rightarrow\mathbbm{C}$, for example, this implies that our normalization must 
differ from the one used e.g.~in~\cite{Keppeler:2012ih} where the same operators are understood as 
projectors $\rep{8}^{\otimes2}\rightarrow\rep{8}^{\otimes2}$. We explain how we obtain correct normalization
for all of our projectors and state the results in appendix~\ref{sec:normalization}. 

\subsection{Construction of the basis invariants}\label{sec:invariants_construction}
To finally obtain the orthogonal basis invariants, the projection operators are contracted with all possible combinations of $H_u$'s and $H_d$'s,
or more precisely, $\boldsymbol{u}^a$'s and $\boldsymbol{d}^a$'s. Using all of the above projection operators, the invariants stated in \eqref{eq:trace_basis_inv} are derived up to 
global numerical, typically $\mathcal{O}(1)$, prefactors (corresponding to the correct normalization of the projection operators) that we drop in~\eqref{eq:trace_basis_inv} for simplicity.

The quadratic traces $I_{20}$, $I_{02}$, and $I_{11}$ are obtained using \eqref{eq:dab} to contract combinations of $\boldsymbol{u}^a$ and $\boldsymbol{d}^a$.
The cubic traces $I_{30}$, $I_{03}$, $I_{21}$ and $I_{12}$ are obtained by contraction with a $d$ tensor invariant projector, eq.~\eqref{eq:cubic}. 
All contractions with the $f$ tensor vanish for symmetry reasons, which we discuss in more detail in the next section. 

All of the quadratic and cubic trace basis invariants are the unique orthogonal invariants at the respective level of contractions in the adjoint space.
This changes at the quartic level. Using the projectors \eqref{eq:P4first} to \eqref{eq:P4last} to derive invariants at the quartic level, multiple different invariants are obtained. Trivially, contractions with \eqref{eq:P4first} yield factorized invariants
of the quadratic level, i.e.~products of $I_{20}$, $I_{02}$, and $I_{11}$. The same is true for non-vanishing invariants that originate from contractions with either all legs being $\boldsymbol{u}^a$ or $\boldsymbol{d}^a$, or with an odd number of $\boldsymbol{u}^a$'s and $\boldsymbol{d}^a$'s (uuud, dddu and permutations). Taking contractions of two $\boldsymbol{u}^a$ and two $\boldsymbol{d}^a$, still, multiple invariants are obtained, in different contraction channels, corresponding to the different operators in eqs.~(\ref{eq:P4first}) to~\eqref{eq:P4last}. Two different non-zero and non-factorized invariants arise from the $\rep{8}_\mathrm{S}$-projection operator, and one invariant each arises from the $\rep{8}_\mathrm{A}$, $\rep{10}$ (an identical one arises from $\overline{\rep{10}}$),
and $\rep{27}$-plet projection operators. The different quartic invariants are, by construction, all orthogonal to each other, and orthogonal to all lower lying invariants, but they are \textit{not} algebraically independent of each other (taking into account the lower lying invariants). In consistency with the Hilbert Series, there is exactly one algebraically independent quartic invariant. 
Amongst all orthogonal quartic invariants found, $I_{22}$ of \eqref{eq:trace_basis_inv} -- as constructed from $\rep{8}_\mathrm{S}$ contracted with $H_u$ and $H_d$ -- minimizes the number of terms in the syzygy~\eqref{eq:syzygy}, 
which is why we have chosen to display this particular invariant and use it in the numerical analysis below. This does not mean that our expression for $I_{22}$
has the fewest possible number of terms when spelled out in matrix elements of $\widetilde{H}_{u,d}$. We have found much shorter (fewer number of terms) orthogonal and algebraically independent quartic invariants, which, however, give rise to more complicated syzygies. The shortest invariant found is the one obtained from $\rep{8}_\mathrm{A}$ which has $189$ terms (as compared to $I_{22}$, which has $294$ terms).

\subsection{CP transformation of the basis invariants}\label{sec:CP_trafo}
Let us now derive general rules for the transformation of projection operators under charge-parity (CP) transformations.
The most general physical CP transformation is given by a simultaneous complex conjugation outer automorphism of all involved symmetry groups~\cite{Trautner:2016ezn}.
Hence, CP also acts as a complex conjugation outer automorphism in flavor space. 
The most general possible CP transformation acts on quark field multiplets as\footnote{%
Here we have suppressed a possibly non-trivial action of the CP transformation in the gauge representation space of each field for the simple reason that it is always possible to rotate 
this transformation to an identity matrix $U=\mathbbm{1}$ for the gauge groups of the SM~\cite{Grimus:1995zi}.}\cite{Gronau:1986xb,Bernabeu:1986fc}
\begin{align}\label{eq:generalizedCP}
Q_\mathrm{L}(t,\boldsymbol{x})~&\mapsto~U_\mathrm{L}\,\mathcal{C}\,Q_\mathrm{L}^*(t,-\boldsymbol{x})\,,&                             \\
u_\mathrm{R}(t,\boldsymbol{x})~&\mapsto~U_{u,\mathrm{R}}\,\mathcal{C}\,u_\mathrm{R}^*(t,-\boldsymbol{x})\,,&      \\
d_\mathrm{R}(t,\boldsymbol{x})~&\mapsto~U_{d,\mathrm{R}}\,\mathcal{C}\,d_\mathrm{R}^*(t,-\boldsymbol{x})\,.&      
\end{align}
Here, $U_\mathrm{L}$, $U_{u,\mathrm{R}}$, and $U_{d,\mathrm{R}}$ are general $3\times3$ unitary matrices acting in flavor space,
while $\mathcal{C}$ is the charge conjugation matrix of fermions given by $\mathcal{C}=\mathrm{i}\gamma_2\gamma_0$ in the chiral Weyl or Dirac basis of gamma matrices. 
Equivalently, the CP transformation can be viewed as acting on the Yukawa coupling matrices $Y_{u,d}$ as
\begin{align}
 Y_u~&\mapsto~U^\mathrm{T}_{\mathrm{L}}\,Y_u^*\,U^*_{u,\mathrm{R}}\,,& \\
 Y_d~&\mapsto~U^\mathrm{T}_{\mathrm{L}}\,Y_d^*\,U^*_{d,\mathrm{R}}\,.&
\end{align}
From this it is straightforward to derive the transformation of the adjoint space vectors
\begin{align}\label{eq:CPtrafo}
 \boldsymbol{u}^a~&\mapsto~-R^{ab}\,\boldsymbol{u}^b\,,& \\\notag
 \boldsymbol{d}^a~&\mapsto~-R^{ab}\,\boldsymbol{d}^b\,,&
\end{align}
where $R$ is the representation matrix of the CP transformation ($\mathbbm{Z}_2$ outer automorphism of $\mathrm{SU}(N)$) in adjoint space, related to $U_\mathrm{L}$ by the consistency condition~\cite{Grimus:1995zi,Holthausen:2012dk}, see also~\cite{Fallbacher:2015rea,Trautner:2016ezn}
\begin{equation}
 U_\mathrm{L}\,\left(-t^{a}\right)^\mathrm{T}\,U_\mathrm{L}^\dagger~=~R^{ab}\,t^b\;.
\end{equation}
For example, in the standard Gell-Mann basis for the $\mathrm{SU}(3)$ generators of the fundamental representation, $U_\mathrm{L}=\mathbbm{1}$ and $R=\mathrm{diag}(-1,+1,-1,-1,+1,-1,+1,-1)$. 

The transformation of $f$ and $d$ tensors under the CP outer automorphism is given by 
\begin{align}
 f^{abc}~&\mapsto~R^{aa'}\,R^{bb'}\,R^{cc'}\,f^{a'b'c'}~=~f^{abc}\,,& \\
 d^{abc}~&\mapsto~R^{aa'}\,R^{bb'}\,R^{cc'}\,d^{a'b'c'}~=~-d^{abc}\,.&
\end{align}
Given this, the CP transformation behavior of invariants can easily be read-off from their respective projection operators, 
presuming that their external legs are contracted with objects that transform like \eqref{eq:CPtrafo}.
The rule for all of our invariants is:
\vspace{0.2cm}
\begin{center}
\text{\emph{An invariant obtained by projection is CP even (CP odd),}} \\
\text{\emph{if the corresponding projection operator contains an even(odd) number of $f$ tensors.}}
\end{center}\vspace{0.2cm}
Without surprise, this shows that the Jarlskog invariant constructed via the operator~\eqref{eq:Jarlskog}
is the only non-vanishing CP-odd invariant in our construction. 

However, note that there is a potential CP-odd invariant lurking already at the cubic level, $k=3$, given by the projection with the operator $\mathrm{i}f^{abc}$.
Only by accident this invariant vanishes in the SM, since there are only two independent tensors, $\boldsymbol{u}^a$ and $\boldsymbol{d}^a$ (from $H_u$ and $H_d$), 
while $f^{abc}$ is totally anti-symmetric. The same argument is true for other potentially CP-odd projections at the levels $k=4$ and $k=5$ (e.g.\ the $\rep{8}$-plet transition operators shown in eq.~\eqref{eq:8sa}). 
The lowest order CP-odd invariant in the SM then arises only at level $k=6$ and is, therefore, highly suppressed.

Note that in more general models, CP violation would generically arise at a lower order. For example, in SM extensions with more than two independent 
structures in the left-handed quark flavor space, CP violation would arise already at the cubic level. This could be the case upon taking into account higher-dimensional operators in effective field theories of the SM~\cite{Bonnefoy:2023bzx} (beyond the paradigm of minimal flavor violation~\cite{DAmbrosio:2002vsn}), or, more concretely, in models with some level of ``quark-lepton unification''. For example, if left-handed charged leptons would, at some scale, be unified with the left-handed quarks. 
In this case, the charged lepton Yukawa couplings $H_\ell:=Y_\ell Y_\ell^\dagger$ would form a third object in the left-handed adjoint flavor space, thereby allowing the CP-odd invariant
\begin{equation}
 \birdtrack{20ex}{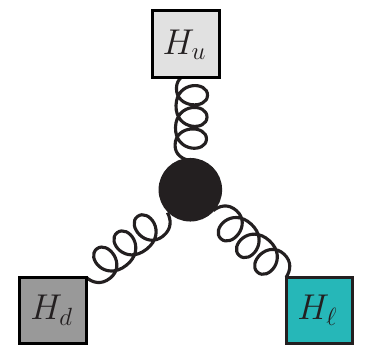}=\mathrm{i}f^{abc}\,\Tr[Y_u^\dagger\,t^a\,Y_u]\Tr[Y_d^\dagger\,t^b\,Y_d]\Tr[Y_\ell^\dagger\,t^c\,Y_\ell]\;.
\end{equation}
This invariant would give rise to CP violation with a strength roughly given by the square-root of the Jarlskog invariant,
i.e.\ lifting the accidental suppression of CPV in the SM by a factor $\mathcal{O}(1/\sqrt{|J|})\sim10^{12}$. 
Studying the details of such an enhancement and its effect on baryogenesis 
is beyond the scope of the present paper. However, it is obvious that such enhancement effects must be taken into account when discussing
the generation of matter-antimatter asymmetry in SM extensions with (left-handed) quark-lepton unification, 
and this might vastly improve our quantitative understanding of Baryogenesis.

\section{Parameter space and experimental values of quark flavor invariants}\label{sec:quantitative}
Having the orthogonal basis invariants at hand, let us quantitatively analyze them in order to obtain a basis invariant picture of the quark flavor puzzle.

On the one hand, we can scan the allowed parameter space (preferentially with a measure that also swipes the corners) to obtain a picture of the landscape of possibilities. 
On the other hand, all physical parameters of the quark sector have been experimentally determined with high accuracy, 
which also experimentally fixes the values of all invariants and their observational uncertainties. 
This is where models \textit{like} the SM (same field content and symmetries but different numerical values of the parameters) 
are differentiated from \textit{the} SM, as determined by observations of Nature. 

\begin{figure}[t]
\begin{minipage}[c]{0.5\textwidth}
\centerline{\includegraphics[width=1.0\textwidth]{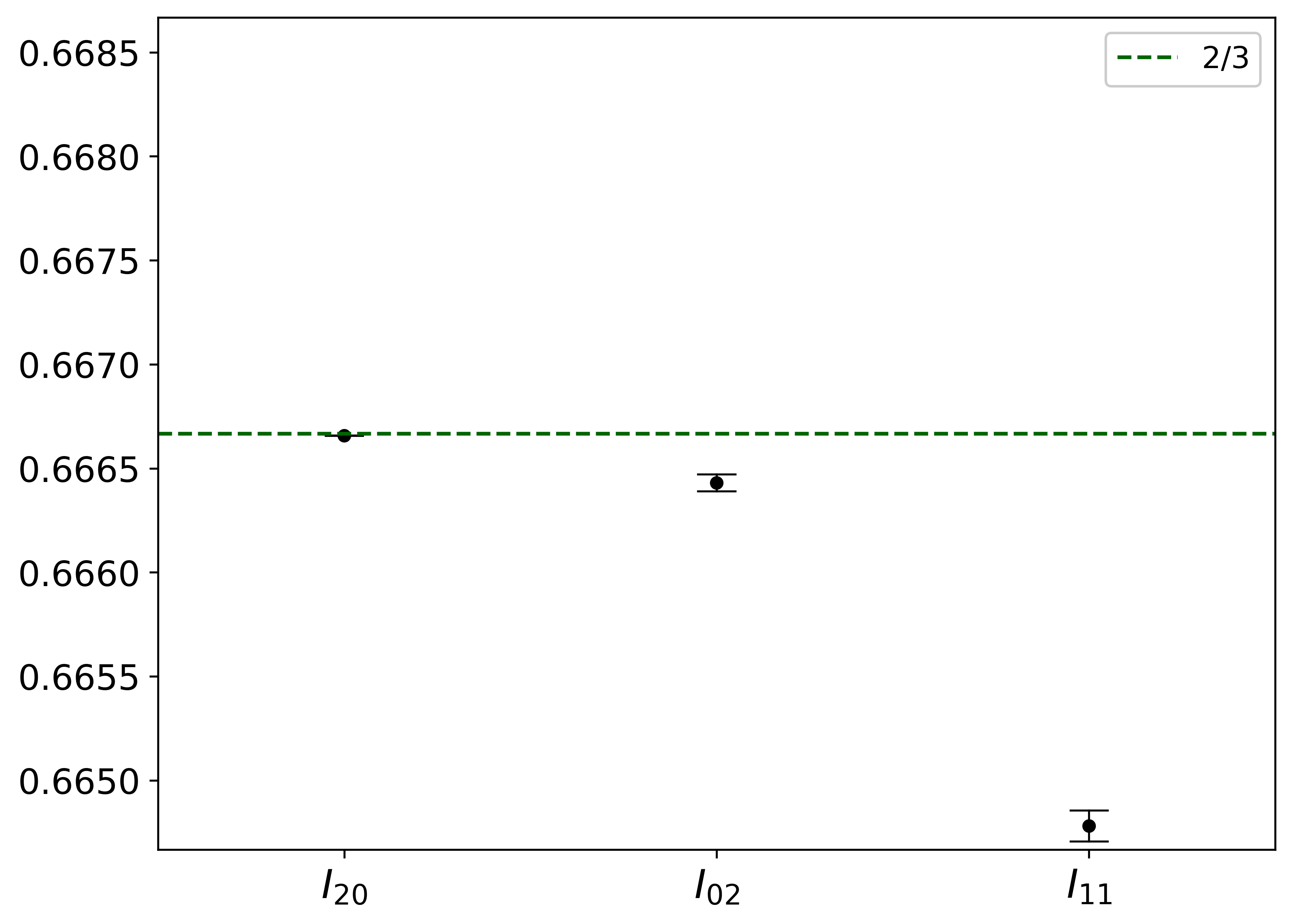}}
\end{minipage}%
\begin{minipage}[c]{0.5\textwidth}
\centerline{\includegraphics[width=1.0\textwidth]{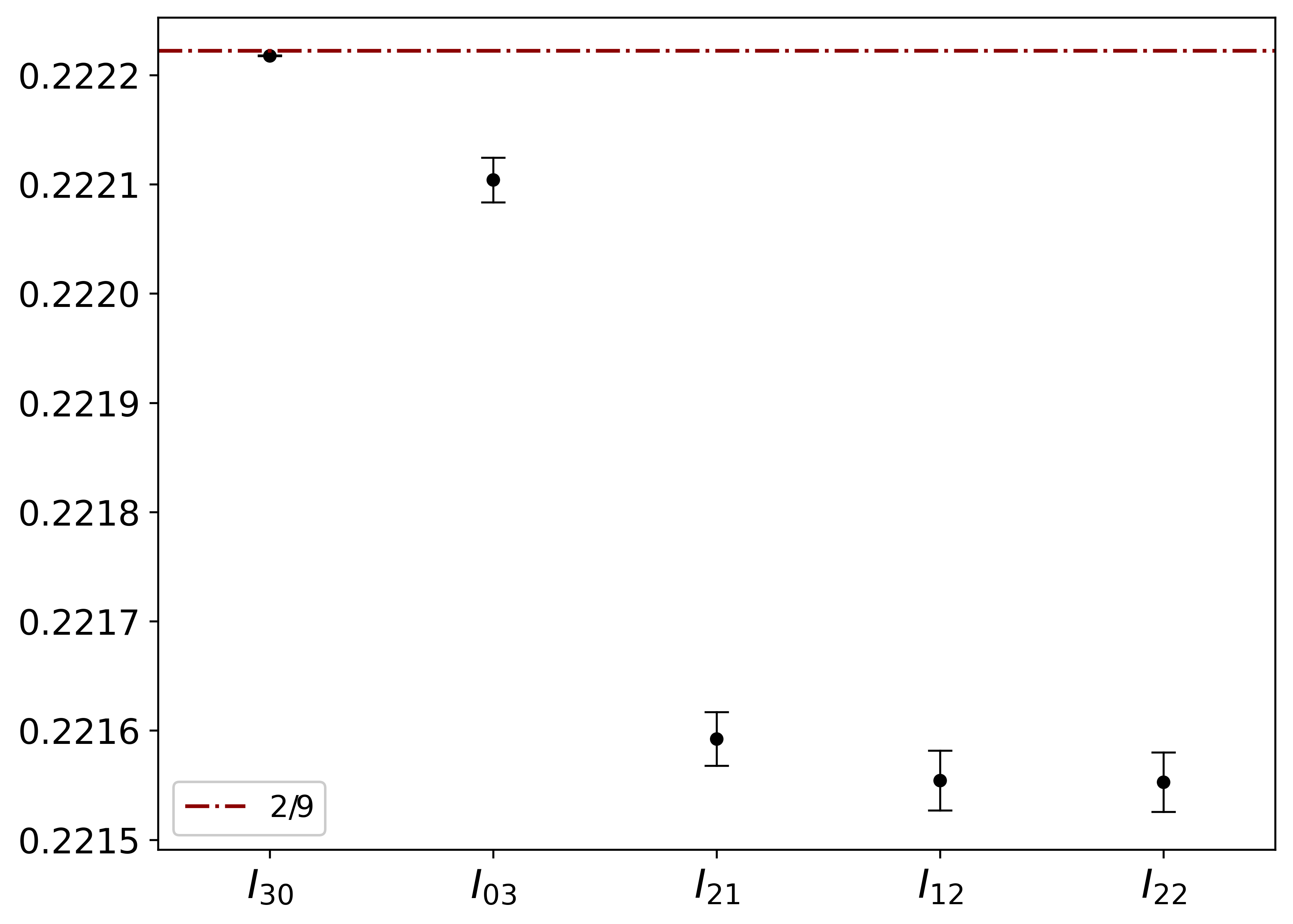}}
\end{minipage}
\caption{\label{fig:cigar}
Experimentally determined values of the orthogonal quark sector basis invariants $\hat{I}_{ij}$, normalized according to~\eqref{eq:normalization}, with $1\sigma$ experimental errors.}
\end{figure}
Using the physical parameters collected in~\cite{ParticleDataGroup:2022pth} for the CKM, and the masses renormalized at the electroweak scale $\mu=M_Z$, see e.g.~\cite{Huang:2020hdv}, the orthogonal invariants and their errors are evaluated in the left column of table~\ref{tab:invariants}. 
Without loss of generality, one can use the standard parametrization and basis choice of eqs.~(\ref{eq:HuPhysical}) and (\ref{eq:HdPhysical}) for this.

To better display the invariants and their correlations, we normalize them to the corresponding power of the two largest Yukawa couplings (which we define here to be $y_t$ and $y_b$ without loss of generality),
\begin{equation}\label{eq:normalization}
 \hat{I}_{ij}:=\frac{I_{ij}}{\left(y_t^{2}\right)^{i} \left(y_b^{2}\right)^{j}}\;.
\end{equation}
The corresponding numerical results are shown in the right column of table~\ref{tab:invariants} and in figure~\ref{fig:cigar}.\footnote{%
We discuss an alternative, arguably even ``more basis invariant'' normalization in appendix~\ref{App:Ialt}.}
\begin{table}
\centering
\begin{tabular}{cl|cl}
    \hline\hline
    Invariant & best fit and error & Normalized invariant & best fit and error \\
    \hline
    $I_{10}$ & $0.9340(83)$                               & $\hat{I}_{10}$ & $1.00001358(^{+85}_{-88})$ \\
    $I_{01}$ & $2.660(49)\times10^{-4}$                   & $\hat{I}_{01}$ & $1.000351(^{+63}_{-71})$ \\ 
    $I_{20}$ & $0.582(10)$                                & $\hat{I}_{20}$ & $0.66665761(^{+59}_{-57})$ \\  
    $I_{02}$ & $4.71(17)\times10^{-8}$                    & $\hat{I}_{02}$ & $0.666432(^{+47}_{-42})$ \\
    $I_{11}$ & $1.651(45)\times10^{-4}$                   & $\hat{I}_{11}$ & $0.664783(^{+91}_{-87})$ \\  
    $I_{30}$ & $0.1811(48)$                               & $\hat{I}_{30}$ & $0.22221769(^{+29}_{-28})$ \\
    $I_{03}$ & $4.18(23)\times10^{-12}$                   & $\hat{I}_{03}$ & $0.222105(^{+24}_{-21})$ \\ 
    $I_{21}$ & $5.14(^{+18}_{-19})\times10^{-5}$          & $\hat{I}_{21}$ & $0.221593(^{+30}_{-29})$ \\ 
    $I_{12}$ & $1.463(^{+65}_{-68})\times10^{-8}$         & $\hat{I}_{12}$ & $0.221555(^{+38}_{-36})$ \\ 
    $I_{22}$ & $1.366(^{+73}_{-76})\times10^{-8}$         & $\hat{I}_{22}$ & $0.221554(^{+38}_{-36})$ \\
    \hline
    $J_{33}$ & $4.47(^{+1.23}_{-1.58})\times10^{-24}$     & $\hat{J}_{33}$ & $2.92(^{+0.74}_{-0.93})\times10^{-13}$ \\
    $J$      & $3.08(^{+0.16}_{-0.19})\times10^{-5}$      && \\
    \hline\hline
\end{tabular} 
\caption{\label{tab:invariants}
Numerical values of the quark flavor sector basis invariants evaluated using experimental data collected by the PDG~\cite{ParticleDataGroup:2022pth}.
The uncertainty intervals are obtained by randomly varying the physical parameters within their $1\sigma$ uncertainty intervals.
The left column displays the orthogonal invariants of eqs.~\eqref{eq:trace_basis_inv}, \eqref{eq:J33}, and \eqref{eq:J}. 
The right column displays the same invariants normalized according to eq.~\eqref{eq:normalization}.}
\end{table}

\begin{figure}[t]
\centerline{\includegraphics[width=1.0\textwidth]{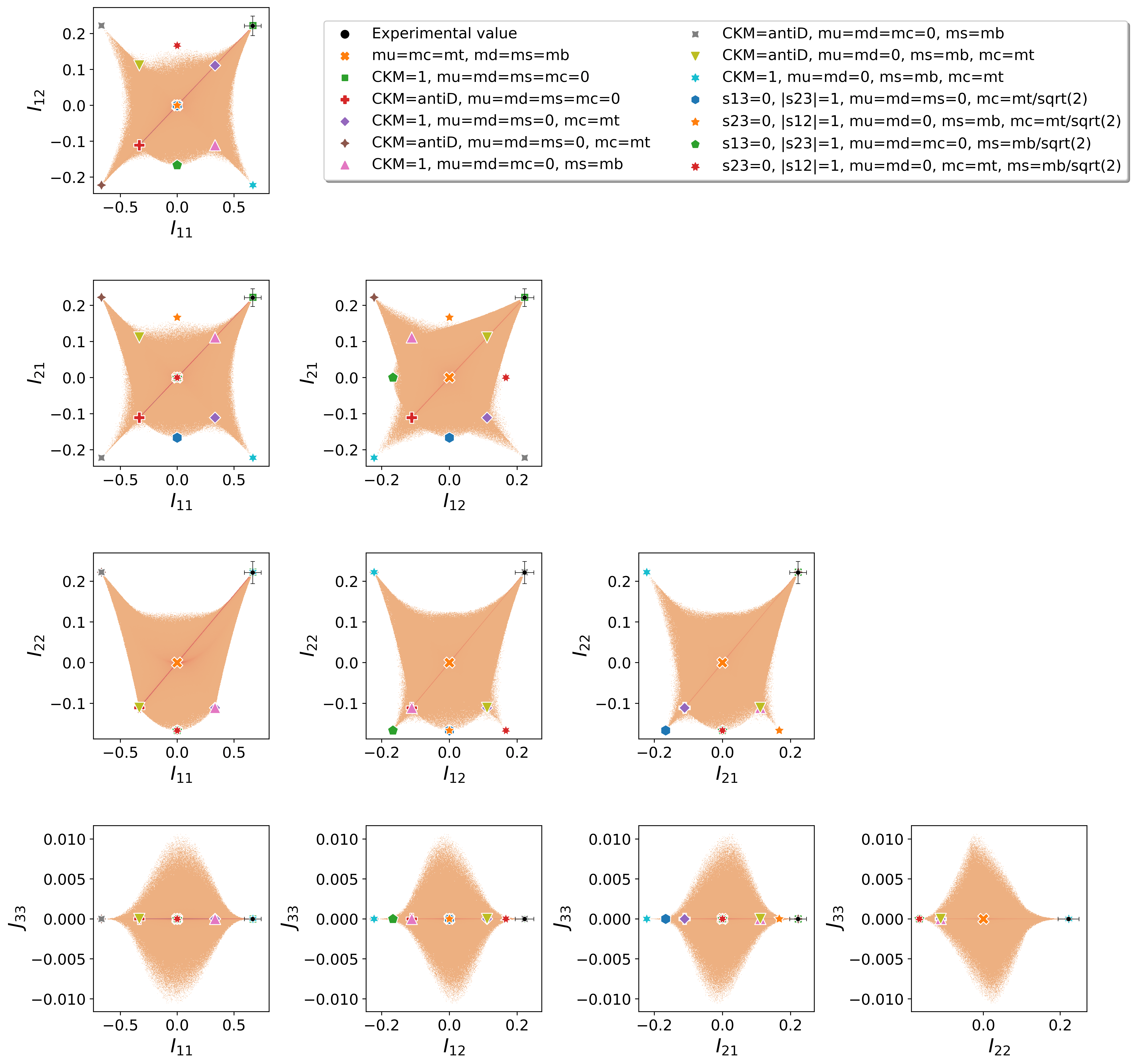}}
  \caption{\label{fig:triangle}
 Parameter space and correlations of orthogonal quark sector flavor basis invariants normalized according to eq.~\eqref{eq:normalization}.
 The experimentally determined values are shown with error bars scaled up by a factor thousand. Points with specific symmetries are marked by symbols according to the legend. Some of the special points overlap for some invariants and we show some more detailed version in figure~\ref{fig:highlight}.}
\end{figure}
\begin{figure}[!h!]
\centerline{\includegraphics[width=1.0\textwidth]{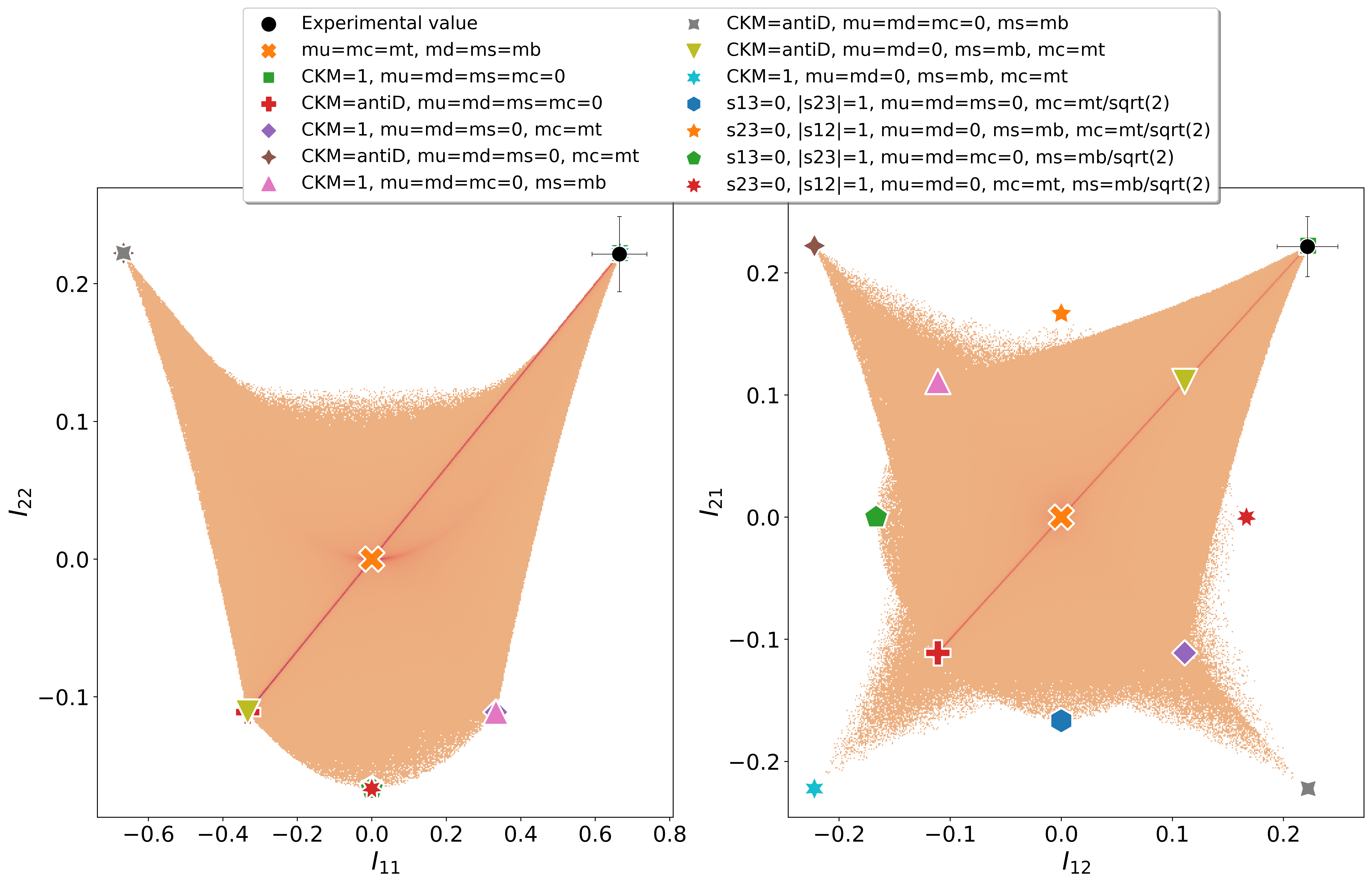}}
  \caption{\label{fig:highlight}
  Parameter space of the invariants $\hat{I}_{11}$, $\hat{I}_{22}$, $\hat{I}_{12}$, and $\hat{I}_{21}$ as determined by two (overlayed) scans over the parameter space with linear and logarithmic measure. The dark diagonal lines originate from the scan with the logarithmic measure which prefers hierarchical physical parameters which lead to strong correlations of the invariants. The special location of the experimentally determined values of the invariants are shown incl.\ their $1\sigma$ error bars scaled up by a factor thousand. We also mark other points in the parameter space corresponding special choices of physical parameters as indicated in the legend ($\text{CKM}=\text{antiD}$ means entries $V_{13}=V_{22}=V_{31}=1$ on the anti-diagonal, and $V_{ij}=0$ everywhere else). 
  On the l.h.s.\ some of the special points overlap according to the rules $\stars{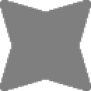}=\stars{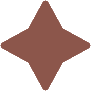}$, $\stars{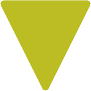}=\stars{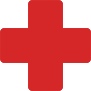}$, $\stars{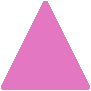}=\stars{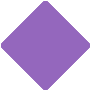}$, $\stars{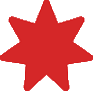}=\stars{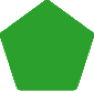}=\stars{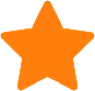}=\stars{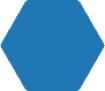}$, $\stars{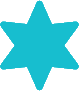}=\stars{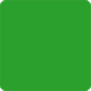}\approx\stars{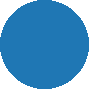}$. On the r.h.s.\ we only have $\stars{1}\approx\stars{3}$.
}
\end{figure}
In order to map out the possible parameter space of the invariants, we perform a scan over the physical parameters.
We scan the parameter space twice, once with a linear measure and once with a logarithmic measure on the physical parameters in 
PDG parametrization. We stress that the goal here is not to explore the likelihood of a given point or even the experimentally observed values, as this would be impossible to determine without knowing the proper measure to be used in the scan. By contrast, we seek to map out \textit{all possible} values of the invariants, i.e.\ the shape of their parameter space when the physical parameters are varied in their physically allowed ranges. 
We use \texttt{NumPy} and evaluate points within a uniform random distribution of CKM angles $s_{12},s_{13},s_{23}\in[-1,1]$ and $\delta\in[-\pi,\pi]$,
as well as masses, either within a uniform random distribution $y_{u,c}\in[0,1]y_t$, $y_{d,s}\in[0,1]y_b$ (``linear'')
or within a uniform random distribution $(m_{u,c}/\mathrm{MeV})\in 10^{[-1,\mathrm{log}(m_t/\mathrm{MeV})]}$, $(m_{d,s}/\mathrm{MeV})\in 10^{[-1,\mathrm{log}(m_b/\mathrm{MeV})]}$ (``logarithmic''). In both cases we only keep points with the ``correct'' mass orderings
$m_{u}<m_{c}$ and $m_{d}<m_{s}$ as not to overweight regions (this is a question of labeling the angles or Yukawa couplings first). Altogether our plots show about $\mathcal{O}(10^7)$ random points. The resulting parameter space of all non-trivial invariants is displayed in figures~\ref{fig:triangle} and~\ref{fig:highlight}. 
The boundedness of the parameter space and the limiting values can be understood by using the Frobenius inner product for our invariants, as explained in detail in appendix~\ref{app:inner_product}.
We highlight several special points in the parameter space as well as the experimentally determined locations of the invariants and their errors (with error bars scaled up by a factor $10^3$ for visibility).

\begin{figure}[t]
\begin{minipage}[c]{0.5\textwidth}
\centerline{\includegraphics[width=1.0\textwidth]{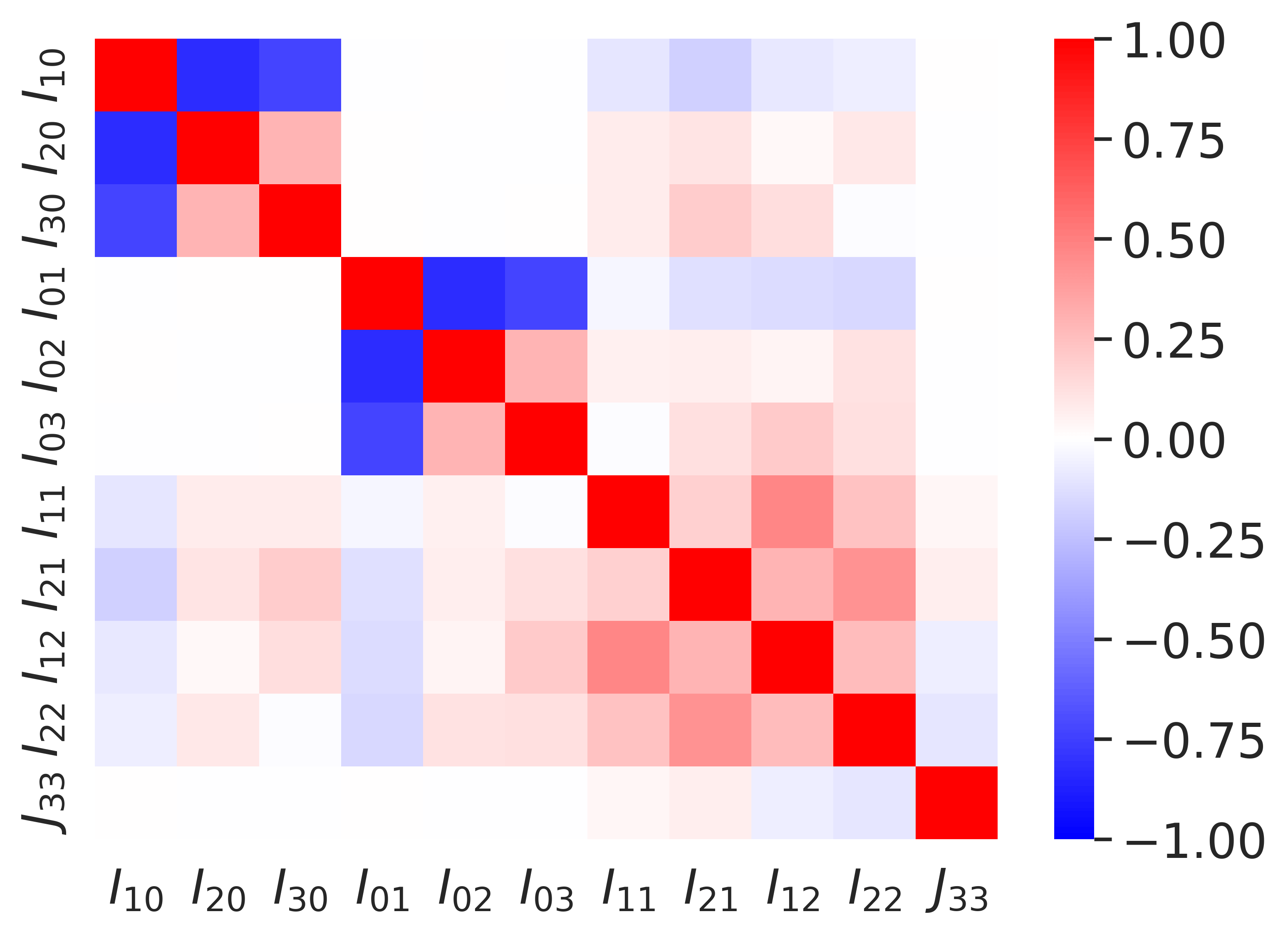}} 
\end{minipage}%
\begin{minipage}[c]{0.5\textwidth}
\centerline{\includegraphics[width=1.0\textwidth]{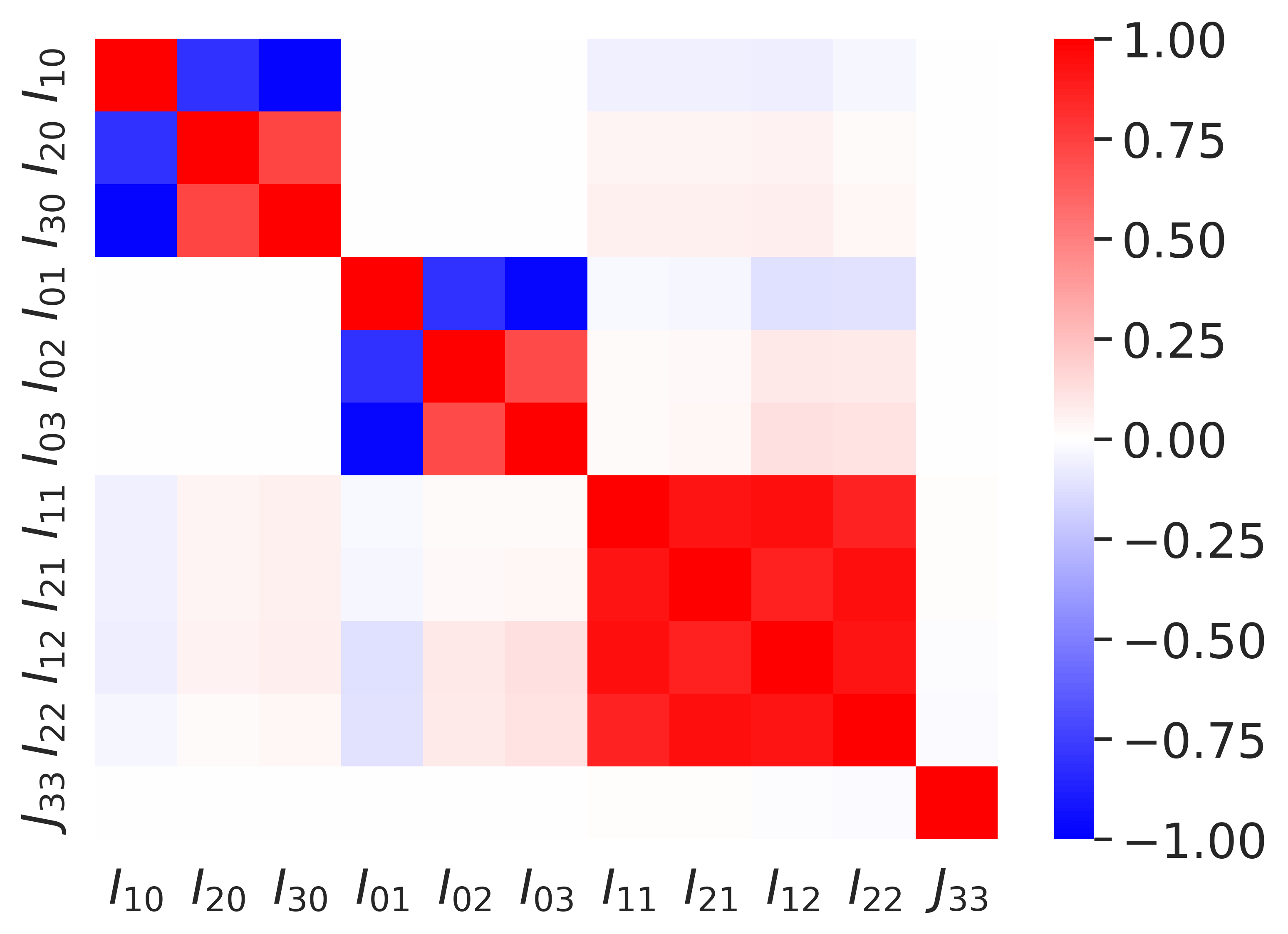}} 
\end{minipage}
\caption{\label{fig:correlations}
  Correlations of all orthogonal basis invariants when scanning over their entire physically allowed parameter space with linear (left) and logarithmic measure (right).}
\end{figure}

Using the physical parameters of the SM as determined from experiment, the following relations turn out to hold approximately
\begin{equation}\label{eq:23}
 \hat{I}_{11}~\approx~\hat{I}_{20}~\approx~\hat{I}_{02}~\lesssim~\frac{2}{3}\;,
\end{equation}
\begin{equation}\label{eq:29}
 \hat{I}_{30}~\approx~\hat{I}_{03}~\approx~\hat{I}_{21}~\approx~\hat{I}_{12}~\approx~\hat{I}_{22}~\lesssim~\frac{2}{9}\;.
\end{equation}
The exact numerical values of the relations should not be over-interpreted as we have used an arbitrary normalization of the invariants
in eq.~\eqref{eq:trace_basis_inv} and not the ``correct'' normalization of the projection operators discussed in appendix~\ref{sec:normalization}.
In particular, the resulting numbers would differ and the approximate equality between cubic and the quartic invariants would not appear. 
Much more interesting is the fact that the experimentally found values of SM parameters realize a situation in which the invariants are sitting very close to their 
maximal possible value, see figure~\ref{fig:triangle}. Furthermore, the invariants display a strong level of positive correlation, see figure~\ref{fig:correlations}.
The correlation between the invariants shown in~\ref{fig:correlations} becomes strong, particularly for the observed (hierarchical) pattern of parameters.
This can also be seen by the darker lines in~\ref{fig:triangle}, corresponding to more points along the diagonal line of positive correlation, arising from the scan of the parameter space with the logarithmic measure that prefers hierarchical parameters. The invariants are not correlated in this way for anarchical patterns of parameters or points in parameter space with otherwise increased flavor symmetry, see the different symbols in figures~\ref{fig:triangle} and~\ref{fig:highlight}.

\enlargethispage{22pt}
The deviations of the normalized invariants from the exact values in eq.~\eqref{eq:23} and \eqref{eq:29} are statistically significant, see table~\ref{tab:invariants} and figure~\ref{fig:cigar}.
The relations become exact in the limit $y_{c,u}\rightarrow0$, $y_{s,d}\rightarrow0$, $\lambda\rightarrow0$ (instead of $\lambda$, also $A\rightarrow0$ is sufficient). Hence, having the invariants fulfill eq.~\eqref{eq:23} and \eqref{eq:29} exactly, 
corresponds to a situation with exact $\mathrm{SU}(2)_{Q_\mathrm{L}}$ flavor symmetry and massless first and second generation quarks, see also appendix~\ref{app:inner_product}. 
Deviations from the exact values of $2/3$ and $2/9$ are given by (highly correlated) leading order negative corrections of size $\mathcal{O}(y_{c,u}^2/y_t^2)$, $\mathcal{O}(y_{s,d}^2/y_b^2)$ and $\mathcal{O}(A^2\lambda^4)$.\footnote{%
\label{foot:flavor}For example, while from table~\ref{tab:invariants} it seems possible that $I_{21}$, $I_{12}$ and $I_{22}$ are actually equal within errors, their errors are highly correlated. The non-vanishing differences of the invariants $\hat{I}_{21}-\hat{I}_{12}\neq0$ and $\hat{I}_{12}-\hat{I}_{22}\neq0$ are statistically significant.} 
The masses of lighter generations, their hierarchies, and deviations from a unity CKM mixing matrix, hence, corresponds to the deviations of the invariants from the symmetric points and the detailed correlation of those deviations. 
Explaining the primary location at the symmetric points and the nature and size of the experimentally significant deviations from the symmetric points 
would amount to solving the flavor puzzle in this language of orthogonal basis invariants. 
 
% In the SM limit of parameters the invariants can approximately be written as Using the SM parametric hierarchy of $y_t\gg y_{c,u}$, $y_b\gg y_{s,d}$ and $\lambda\ll 1$, 
% the normalized invariants to leading order are given by
% \begin{align}
%  \hat{I}_{20}~&=~\frac23-2\frac{y_c^2+y_u^2}{y_t^2}+\mathrm{h.o.}\;,& \\
%  \hat{I}_{02}~&=~\frac23-2\frac{y_s^2+y_d^2}{y_b^2}+\mathrm{h.o.}\;,& \\
%  \hat{I}_{30}~&=~\frac29-\frac{y_c^2+y_u^2}{y_t^2}+\mathrm{h.o.}\;,& \\
%  \hat{I}_{03}~&=~\frac29-\frac{y_s^2+y_d^2}{y_b^2}+\mathrm{h.o.}\;,& \\
%  \hat{I}_{11}~&=~\frac23-A^2\lambda^4-\frac{y_c^2+y_u^2}{y_t^2}-\frac{y_s^2+y_d^2}{y_b^2}+\mathrm{h.o.}\;,& \\ 
% 3\,\hat{I}_{21}~&=~\frac23-A^2\lambda^4-2\frac{y_c^2+y_u^2}{y_t^2}-\frac{y_s^2+y_d^2}{y_b^2}+\mathrm{h.o.}\;,& \\ 
% 3\,\hat{I}_{12}~&=~\frac23-A^2\lambda^4-\frac{y_c^2+y_u^2}{y_t^2}-2\frac{y_s^2+y_d^2}{y_b^2}+\mathrm{h.o.}\;,& \\ 
% 3\,\hat{I}_{22}~&=~\frac23-A^2\lambda^4-2\frac{y_c^2+y_u^2}{y_t^2}-2\frac{y_s^2+y_d^2}{y_b^2}+\mathrm{h.o.}\;.&
% \end{align}
% $\mathrm{h.o.}$ here refers to higher order corrections in $\lambda$ or higher powers of the Yukawa coupling ratios. 
% This shows that relations \eqref{eq:23} and \eqref{eq:29} become exact, in the limit of zero mixing and zero 1st and 2nd-generation fermion masses.

\section{Renormalization group evolution of the orthogonal invariants}\label{sec:renormalization}
\begin{figure}[t]
\begin{minipage}[c]{0.5\textwidth}
\centerline{\includegraphics[width=1.0\textwidth]{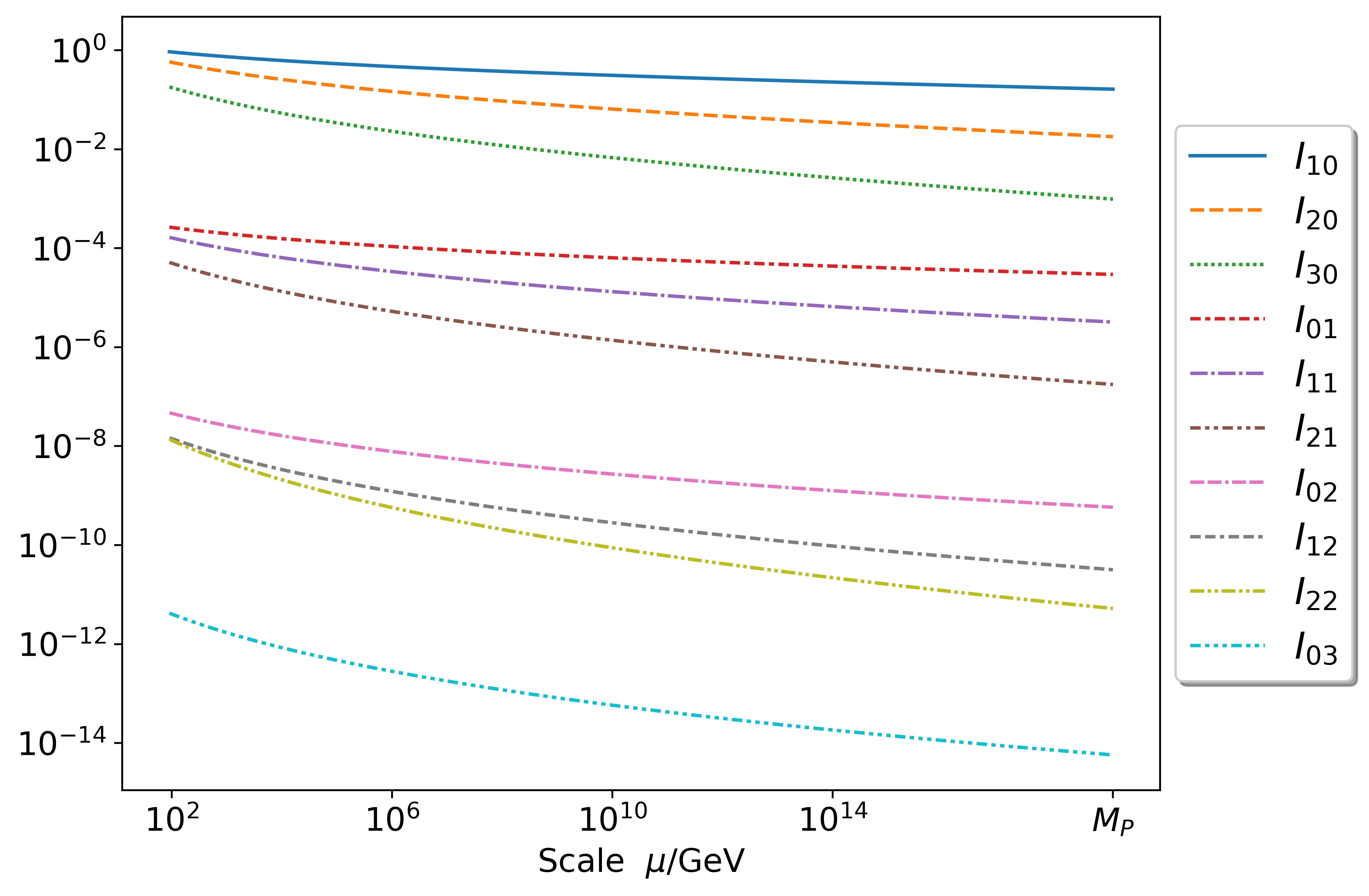}}
\end{minipage}%
\begin{minipage}[c]{0.5\textwidth}
\centerline{\includegraphics[width=1.0\textwidth]{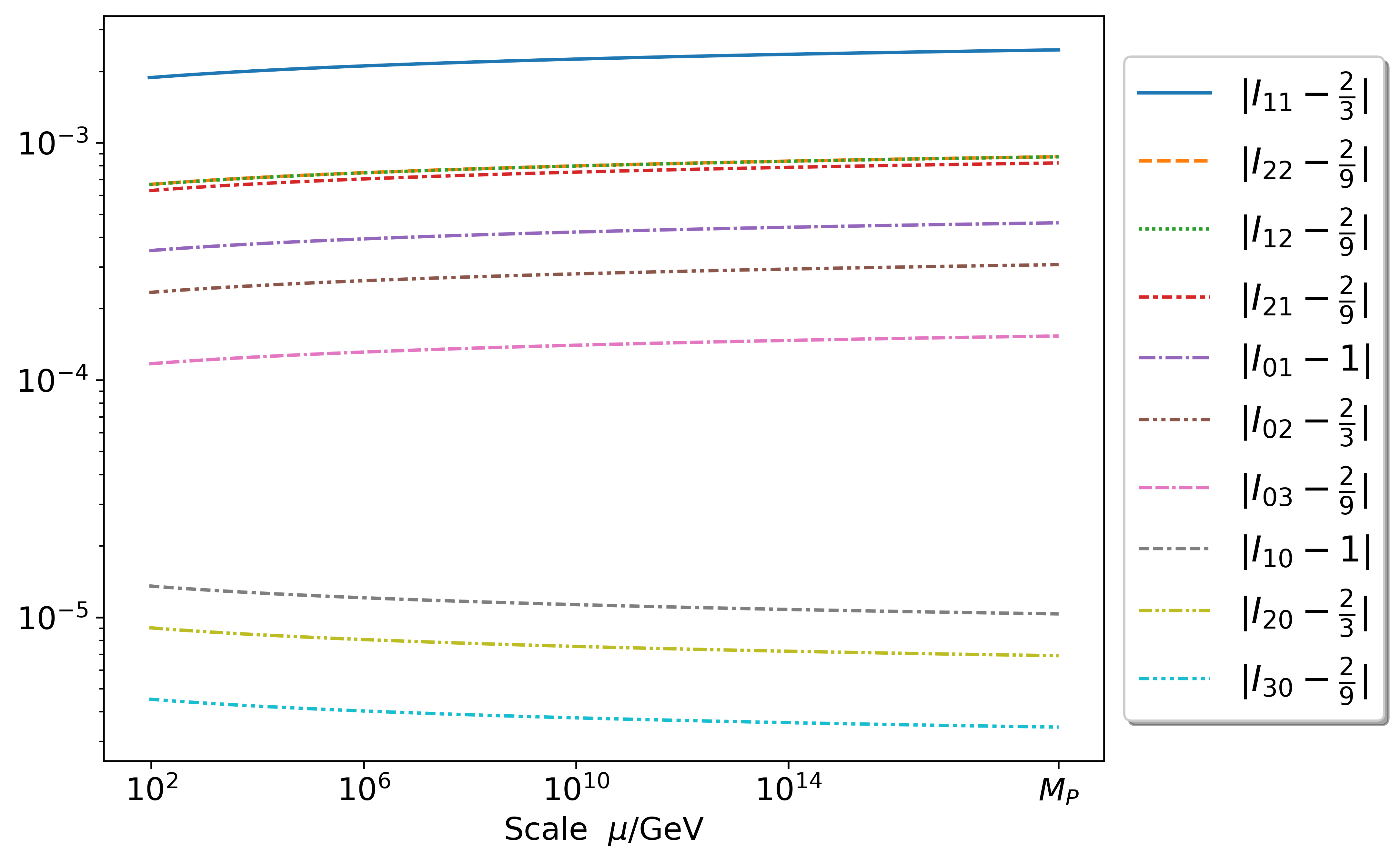}}
\end{minipage}
\caption{\label{fig:RGEs}
Renormalization group running (at one-loop accuracy) of the orthogonal quark sector basis invariants (left) and invariants normalized according to~\eqref{eq:normalization} (right).
}
\end{figure}

To display the evolution of the invariants under the renormalization group we use the one-loop renormalization group equations (RGEs) of~\cite{Ferreira:2010xe}
adopted to our case (see also~\cite{Machacek:1983fi,Sasaki:1986jv,Babu:1987im,Lindner:2005as,Bednyakov:2014pia}). We use the definitions 
\begin{align}
 \mathcal{D}&:=16\pi^2\mu\frac{\mathrm{d}}{\mathrm{d}\mu}\;,& \\
 a_\Delta&:=-8\,g_s^2-\frac94g^2-\frac{17}{12}g'^2\;,& \\
 a_\Gamma&:=-8\,g_s^2-\frac94g^2-\frac5{12}g'^2\;,& \\
 a_\Pi&:=-\frac94 g^2 - \frac{15}{4} g'^2\;,& \\
 t_{udl}&:=3\,\mathrm{Tr}\widetilde{H}_u+3\,\mathrm{Tr}\widetilde{H}_d + \mathrm{Tr}\widetilde{H}_\ell\;,
\end{align}
with $g_s$, $g$, and $g'$ being the respective gauge couplings of $\mathrm{SU}(3)_\mathrm{c}$, $\mathrm{SU}(2)_\mathrm{L}$, and $U(1)_\mathrm{Y}$, normalized such that 
the Higgs doublet has hypercharge $1/2$. The RGEs for $\widetilde{H}_u$ and $\widetilde{H}_d$ are then given by
\begin{align}
 \mathcal{D}\widetilde{H}_u &= 2\left(a_\Delta+t_{udl}\right)\,\widetilde{H}_u + 3\,\widetilde{H}_u^2-\frac{3}{2} \left( \widetilde{H}_d \widetilde{H}_u + \widetilde{H}_u \widetilde{H}_d \right)\;,& \\
 \mathcal{D}\widetilde{H}_d &= 2\left(a_\Gamma+t_{udl}\right)\,\widetilde{H}_{d} + 3\,\widetilde{H}_d^2-\frac{3}{2} \left( \widetilde{H}_d \widetilde{H}_u + \widetilde{H}_u \widetilde{H}_d \right)\;,& \\
 \mathcal{D}\widetilde{H}_\ell &= 2\left(a_\Pi+t_{udl}\right)\,\widetilde{H}_\ell + 3\,\widetilde{H}_\ell^2\;,& 
\end{align}
while the RGEs of the gauge couplings take the standard form
\begin{align}
\mathcal{D}g_s&=-7\,g_s^3\;,& \mathcal{D}g&=-\frac{19}{6}g^3\;,& \mathcal{D}g'&=\frac{41}{6}g'^3\;.& 
\end{align} 
After solving the RGEs for $\widetilde{H}_u$, $\widetilde{H}_d$, and $\widetilde{H}_\ell$, it is straightforward to evaluate the orthogonal invariants 
defined in eqs.~\eqref{eq:trivialInvariants}, \eqref{eq:trace_basis_inv}, and~\eqref{eq:Jdefinition} at any scale. 
We show the result for the invariants in figure~\ref{fig:RGEs} (left) and for the normalized invariants in figure~\ref{fig:RGEs} (right).
No particularly striking feature is happening in the RGE evolution from the electroweak scale up to very high scales, and we display the running up to the Planck scale. At a scale $\mu\sim10^{41}\,\mathrm{GeV}$ the invariants turn zero, while running to lower scales the invariants become infinite at a scale $\mu\sim{40}\,\mathrm{MeV}$ (some of the invariants, but not their normalized counterparts, show crossings in the RGE flow at scales below $\sim10\,\mathrm{GeV}$). We do not consider threshold effects and matching here, but note that integrating out fermions unavoidably corresponds to changing the ring and its structure. Hence, the significance of these scales should be evaluated using a precise evaluation of higher-loop order RGEs and proper treatment of thresholds below the electroweak scale, which is beyond the scope of this work.

As an important crosscheck, we explicitly confirm that we can use our scale dependent invariants to extract the running of the masses 
in agreement with the results of~\cite[Table~2]{Huang:2020hdv} (within reasonable errors to be blamed on one vs.\ three-loop accuracy).
We also confirm the correct running of CKM elements, Wolfenstein parameters, and $J$ compared to~\cite{Babu:1987im,Grossman:2022ehc}, as discussed in detail in appendix~\ref{app:CKMrunning}.

Even though the normalized invariants do evolve very little, see r.h.s.\ of fig.~\ref{fig:RGEs},
their evolution is significant as compared to the error budget of the invariants at the electroweak scale (see table~\ref{tab:invariants}, right).
For a direct construction of RGE evolution invariants at the one loop order we refer to~\cite{Harrison:2010mt,Feldmann:2015nia}.
Using the orthogonal invariants and their directly derived RGE equations (to be presented elsewhere) will also enable the future construction of higher-loop-order RGE invariant expressions, or allow to show that they do not exist. Formulating the running of invariants directly and exactly in terms of the invariants themselves is a formidable task for future work.

\newpage
\section{Discussion and Comments}\label{sec:discussion}
Let us give some remarks about the construction of the invariants, numerical results and directions for future work.

\begin{itemize}
 \item We have constructed our invariants such that they are orthogonal in adjoint space of left-handed quark flavor, $\mathrm{SU}(3)_{Q_\mathrm{L}}$. We are aware of at least one other distinct orthogonal basis to construct the SM flavor basis invariants,
 namely from orthogonal projectors in the fundamental space of $\mathrm{SU}(3)_{Q_\mathrm{L}}$. Depending on the application, it may be more appropriate to work with one orthogonal basis for the invariants or the other.
 The construction of the orthogonal basis in fundamental space requires the construction of orthogonal projection operators up to $(\rep{\bar{3}}\otimes\rep{3})^{\otimes6}\rightarrow\mathbbm{C}$.
 Those can be constructed via Young tableaux (pulling up or down one of the \mbox{(anti-)fundamental} indices), see~\cite{Trautner:2018ipq}, and require several steps of \mbox{(anti-)symmetrization} of up to $18$ fundamental
 indices of $\mathrm{SU}(3)$. On the one hand, such complicated projection operators can straightforwardly be constructed by hand using birdtrack technology which nicely generalizes, for example to~$\mathrm{SU}(N)$.
 On the other hand, evaluating the according operators explicitly is an intensive task of high complexity that easily exhausts memory capacities even of large computing clusters
 (this is a problem with large number of possible permutations and the computational effort grows roughly proportional to the factorial of the number of indices). 
 Hence, eventually this task should be delegated to super- or even quantum computers (and quantum analogue simulators) where the construction of invariant operators may even serve as useful benchmark problem. Once the operators are constructed, they are small in memory size and their correctness is computationally cheap to confirm.
 
 \item In the adjoint space construction of orthogonal basis invariants, there is an ambiguity in choosing $I_{22}$ (see discussion in sec.~\ref{sec:invariants_construction}). On the one hand, there could be orthogonal bases other than in the adjoint space in which there exist a unique orthogonal quartic invariant $I_{22}$. On the other hand, the origin of distinct orthogonal quartic invariants from different covariant contraction channels 
 in the adjoint space could be important to understand the \mbox{(mis-)alignment} of the $\rep{8}$-plet vectors $\boldsymbol{u}^{a}$ and $\boldsymbol{d}^{a}$, which is instrumental for the detection of flavor symmetries and 
 might be very relevant in order to understand the observed flavor structure of the SM. Depending on the application it may be more appropriate to work with one or the other choice of quartic invariant, or even with multiple of them simultaneously. 
 
 \item Using orthogonal projection operators automatically tracks the origin of the invariants from specific contraction of covariants.
 Specific alignment of covariants is in one-to-one relation with a corresponding relation between the basis invariants~\cite{Merle:2011vy,Bento:2020jei}.
 The importance of the relative alignment of basis covariant quantities for the detection of flavor symmetries is known from 2HDM~\cite{Ivanov:2005hg, Nishi:2006tg, Maniatis:2006fs, Maniatis:2007vn, Ferreira:2010yh} as well as 3HDM~\cite{Ivanov:2018ime,Ivanov:2019kyh,deMedeirosVarzielas:2019rrp}. It is clear that the alignment of covariants will also play an important role in classifying and detecting all possible flavor symmetries in the parameter space of the SM.
 
 \item Thinking about the alignment of covariants in the SM also leads the way to investigate symmetries of the invariants under $u\leftrightarrow d$ exchange.\footnote{
 This does not necessarily have to be a discrete transformation but could also be a continuous rotation -- a three-generation generalization of the flavor isospin $\mathrm{SU}(2)$.}
 Interestingly such ``custodial flavor'' transformations are usually not considered as flavor symmetries in the sense that they are not subgroups of $\mathrm{SU}(3)_{Q_\mathrm{L}}\otimes\mathrm{SU}(3)_{u_\mathrm{R}}\otimes\mathrm{SU}(3)_{d_\mathrm{R}}$ (under which the basis invariants are, by construction, invariant). Note that the whole HS and PL construction of sec.~\ref{sec:construction} is automatically symmetric under the exchange of $u\leftrightarrow d$, simply because up and down sectors transform in identical representations under $\mathrm{SU}(3)_{Q_\mathrm{L}}$ flavor basis changes. By contrast, the basis invariants do generally not obey the $u\leftrightarrow d$ exchange symmetry and this is specifically the case for the experimentally determined invariants, see tab.~\ref{tab:invariants}.\footnote{%
 Requesting exact $u\leftrightarrow d$ exchange symmetry would imply $Y_u=Y_d$, hence $H_u=H_d$, and consequently $I_{01}=I_{10}$, $I_{02}=I_{20}=I_{11}$, $I_{03}=I_{30}=I_{21}=I_{12}$, factorization of $I_{22}$ into smaller invariants, and $I_{33}=0$, i.e.\ absence of CP violation. Note that $u\leftrightarrow d$ symmetry would \textit{not} automatically imply saturation of the inequalities eq.~\eqref{eq:23} and~\eqref{eq:29}, see discussion in app.~\ref{app:inner_product}.} Hence, the behavior of invariants under permutation of up- and down-sector structures is an analysis tool to detect flavor symmetry and ``custodial flavor'' symmetry violation beyond the usually considered subgroups of $\mathrm{SU}(3)_{Q_\mathrm{L}}\otimes\mathrm{SU}(3)_{u_\mathrm{R}}\otimes\mathrm{SU}(3)_{d_\mathrm{R}}$. 
 In this respect, we note that our choice of orthogonal invariants are either symmetric under up- and down-sector exchange or are simply being pairwisely permuted (which allows to form $u\leftrightarrow d$ symmetric and anti-symmetric combinations). 
 While the various quartic invariants are all $u\leftrightarrow d$ even, the $u\leftrightarrow d$ anti-symmetric combinations of invariants such as $\hat{I}_{21}-\hat{I}_{12}$ seem to be particularly relevant to 
 explore the custodial flavor symmetry breaking (see also the discussion in footnote~\ref{foot:flavor}). Also the Jarlskog invariant is odd under $u\leftrightarrow d$, hence, even under the combined transformation of CP and $u\leftrightarrow d$, providing a link of CP and flavor transformations that shall also be further studied. 
 
 \item The close to maximal correlation of all invariants correspond to close-to minimization of the absolute value of the Jarlskog invariant, see fig.~\ref{fig:triangle}.
 That is, CP violation in the SM as measured by the absolute value of $J_{33}$ is much smaller than it could be (which is, of course, well known) but in the invariant language it is evident that a larger positive value and high correlation of the CP-even invariants (which itself corresponds to large parameter hierarchies) corresponds to less CP violation. It remains to be seen whether this can help to explain the observed structure.

 \item It is clear that any successful approach to the flavor puzzle must explain the special values of the invariants close to their maximum values, including the small but significant deviations from the maximal values, as well as the strong correlation of the invariants. Being aware of the special locations of the orthogonal invariants and their explicit covariant content may help in order to resolve the flavor puzzle. It remains to be seen whether the necessary alignment and the resulting parameters can be explained in a conventional QFT and model building approaches with spurion potentials, see e.g.~\cite{Feldmann:2009dc, Grinstein:2010ve, Nardi:2011st, Alonso:2011yg, Espinosa:2012uu, Fong:2013dnk,Altshuler:2023eaw}, or otherwise, for example using radiative corrections (see \cite{Weinberg:2020zba} and references therein), textures (see e.g.~\cite{Xing:2020ijf}), discrete or modular flavor symmetries (see e.g.~\cite{King:2013eh, Feruglio:2015jfa, Feruglio:2017spp,Kobayashi:2023zzc} for reviews), or more exotic approaches, such as explaining the parameters of Nature by entanglement or entropy arguments~\cite{Jaynes:1957zza,Bousso:2007kq,dEnterria:2012eip,Alves:2014ksa,Beane:2018oxh, Low:2021ufv, Quinta:2022sgq,Miller:2023snw,Carena:2023vjc}. In fact, the latter approach seems to be particularly attractive here, as it is feasible that the maximum correlation of orthogonal invariants corresponds to a stationary point of quantum information theoretic von Neumann or Shannon entropy.
 
 \item Physical observables must not depend on an unphysical choice of basis and parametrization.
 Hence, all physical observables must be expressible in terms of basis invariant quantities.
 For the SM, and SM+4th generation rephasing invariants this was already discussed in~\cite{Branco:1987mj,Bjorken:1987tr},
 while for typical extensions like the 2HDM it was discussed in~\cite{Lebedev:2002wq,Davidson:2005cw,Haber:2006ue}, and more recently~\cite{Grzadkowski:2016szj,Ogreid:2018bjq,Boto:2020wyf,Ferreira:2022gjh}.

 However, presently it is unclear how basis invariants can, in general, be related to all possible
 physical observables of a theory, and this should also be clarified.
 An important observation in this context is that basis invariants always correspond to closed-form diagrams
 akin to ``vacuum bubble'' Feynman diagrams (for an example, see e.g.~\cite[Figs.~3 and 4]{Nilles:2018wex}). Hence, a conjecture is that the general relation 
 between basis invariants and physical observables can be made via the well-known optical theorem, in the spirit of modern amplitude methods.
 Here, bubble diagrams correspond to forward-scattering vacuum-to-vacuum transitions and it shall be explicitly explored whether 
 successive Cutkosky cuts~\cite{Cutkosky:1960sp} of vacuum bubble diagrams are sufficient to relate the basis invariants of a model to all physical observables 
 such as cross sections and decay rates. Aspects of this technique have been pioneered in~\cite{Jarlskog:1987zd,Botella:1994cs,Botella:2004ks, Blazek:2021zoj} and should be reanalyzed
 with respect to orthogonal basis invariants and finally merged with modern amplitude methods.

 \item Regarding the running of basis invariants and the derivation of RGE invariants, it seems most promising to use conformal transformation of the invariants in order to derive their RGEs directly. By contrast, trying to anticipate the RGEs of invariants from \textit{truncated} perturbative treatment of running of physical parameters (which is trustworthy for the physical parameters) may lead to RGEs of invariants that are \textit{not} trustworthy (because they would automatically involve higher powers of the couplings not covered by the RGE expansion for the physical parameters). The power counting in terms of invariants is different than the power counting in terms of the physical parameters. 
 An important crosscheck to pass for any system of RGEs of invariants, is that these do indeed reproduce the $n$-loop running of physical parameters used to derive them, and we emphasize that RGEs of invariants that do not pass this crosscheck are not trustworthy.
 
 \item Finally, we re-iterate that our method is intrinsically non-perturbative. 
 It may be interesting to think about the invariants at the QCD scale.
 Our invariants are \textit{exact} to all orders in the Yukawa couplings, implying that the couplings could be arbitrarily large and we could still evaluate the invariants exactly. 
 This means that our invariants are also exact in any kind of light or heavy flavor expansion, or working in specific limits like isospin symmetry.  Nonetheless, using such approximations may of course be helpful in analyzing the invariants, or in expressing experimental observables as functions of the invariants in their associated symmetric limits. Since all observables ought, in principle, to be basis invariant and our computations here should also facilitate derivation of basis invariant expressions for otherwise perturbative expressions. A formidable first task of this kind would be to explicitly show the proportionality of all measured CP violating observables in the SM to the Jarlskog invariant.
 
 \end{itemize}

\section{Conclusions}\label{sec:conclusions}
This paper provides the first quantitative, entirely basis independent characterization of the Standard Model quark flavor puzzle.
To achieve this, we have explicitly constructed an orthogonal basis for the ring of flavor basis invariants, using hermitian projection operators derived via birdtrack diagrams in adjoint flavor space. The virtue of constructing the invariants using orthogonal hermitian projection operators is that the invariants are as short as possible by construction and their covariant content, and transformation behavior under symmetries and CP, is explicit. Furthermore, the orthogonal invariants give rise to the shortest syzygy known to date, which relates
the CP-odd Jarlskog invariant to the set of ten CP-even primary invariants. These ten primary invariants correspond to the ten well known physical parameters of the quark sector, 
but provide an intrinsically non-perturbative view on the parameter space. 

We have explored the full parameter space of the invariants with a scan, and firstly ``measured'' the value of the orthogonal invariants as determined by experiments. At the parameter point realized by Nature, we find the invariants are close to maximally correlated and assume close to maximal values (besides the Jarlskog invariant, which is close to minimized in absolute value by the fact that the other invariants are maximized). The deviations from the maximal possible values of the invariants correspond to the subleading parameters of the quark flavor sector, i.e.\ light Yukawa couplings and small mixing angles. We have also investigated the renormalization group evolution of the invariants and find that the appropriately normalized invariants are close to RGE invariant up to scales much higher than the Planck mass. 

Alongside the main line of the paper, we have given comments about other possible orthogonal bases for the flavor invariants, the correct absolute normalization of the invariants, accidentally vanishing order-3 CP violation in the SM, detection of symmetries using covariant alignment and invariant relations, as well as about the general relation of basis invariants to observables.

The quark flavor puzzle in invariants may be phrased as: Why are the invariants so strongly correlated, and what explains their tiny deviation from the maximal possible values?
We hope that our treatment provides clarity and guidance for model building, to ultimately describe and understand the flavor puzzle with fewer parameters than in the~SM.

\section*{Acknowledgments}
We would like to thank Renato Fonseca for discussions on related projects. AT is grateful to Claudia Hagedorn for discussions on related projects and to Maximilian Berbig for an insightful observation. All birdtrack diagrams of this work have been generated with \texttt{JaxoDraw}~\cite{jaxodraw}.

\section*{Notes added}
After the initial publication of our work, it was kindly pointed out to us by Bingrong Yu that a very similar set of generators for the ring of invariants of two $3\times3$ matrices 
was derived earlier, in the mathematics literature and without physical context, see~\cite{Teranishi:1986,Aslaksen:2006}. The invariants of traceless matrices used by~\cite{Aslaksen:2006} 
coincide with ours defined in~\eqref{eq:trace_basis_inv} and~\eqref{eq:Jdefinition} (besides $I_{22}$ and $J_{33}$), however, derived without a notion of (projection operator) orthogonality. The invariant combination $v$ defined in \cite{Aslaksen:2006} corresponds to the quartic invariant $I'_{22}:=\Tr(H_u^2H_d^2)-\Tr(H_uH_dH_uH_d)$ constructed in our language by using the projection operator $\rep{8}_\mathrm{A}$, see discussion in sec.~\ref{sec:construction} (their $w$ corresponds to $J_{33}$).
Furthermore, it was pointed out to us by the referee that the multi-graded Hilbert series~\eqref{eq:hilbert_graded} for a pair of complex $3\times3$ matrices, as well as generalizations thereof to a higher number of generations, 
were already computed in~\cite{Djokovic:2006}.

\appendix
\section{The Hilbert series of a CP conserving theory}\label{app:CP-even_ring}
Here we show a nice way to describe a CP conserving subring of the SM. For this, note that under the special subgroup of $\mathrm{SU}(3)\supset\mathrm{SO}(3)$, the branching of the adjoint is $\rep{8}\rightarrow\rep{5}\oplus\rep{3}$.
Note that under the CP outer automorphism, $\rep{8}$-plets transform as \eqref{eq:CPtrafo} implying that for the $\mathrm{SO}(3)$ irreps: 
\begin{equation}
 \mathrm{CP:} \quad \rep{5}\mapsto\rep{5}\;, \qquad\text{and}\qquad \rep{3}\mapsto-\rep{3}\;.
\end{equation}
Hence, a sufficient condition to obtain a CP even ring is to set the triplets to zero $\rep{3}\rightarrow0$.\footnote{%
We denote the absence of CP-odd triplet covariants here only as sufficient (not also necessary) condition, noting that even powers of CP-odd triplets may form CP-even invariants. 
}
This turns the hermitian matrices $\widetilde{H}_{u,d}$ to symmetric matrices. The group acting on the five-plets now is not the full $\mathrm{SU}(3)$ but only the $\mathrm{SO}(3)$ subgroup. Due to the well-known isomorphism of the Lie algebras $\mathfrak{so}(3) \cong \mathfrak{su}(2)$, we can write the HS for this ring as HS of $\mathrm{SU}(2)$ without loss of generality. The two trivial singlets $\mathrm{Tr}(\widetilde{H}_{u,d})$ stay exactly the same. 

The HS is computed as (we use dummy indices $u$ and $d$ here but this time for the $\rep{5}$-plets alone)
\begin{equation}
    H(K[V]^G; u, d) = \int_{\mathrm{SU}(2)} d \mu_{\mathrm{SU}(2)} \,
    \mathrm{PE}\left[ z_1, z_2 ; u; \mathbf{5} \right]
    \mathrm{PE}\left[ z_1, z_2 ; d; \mathbf{5} \right] \,.
\end{equation}
This yields
\begin{equation}
    H(K[V]^G; u, d) = \frac{1 + u^2 d^2 + u^4 d^4}
    {(1 - u^2) (1 - d^2) (1 - ud) (1 - u^3) (1 - d^3)
    (1 - ud^2) (1 - u^2 d)} \, ,
\end{equation}
and its ungraded version ($t=u=d$)
\begin{equation}
    H(K[V]^G; t) = \frac{1 + t^4 + t^8}
    {(1 - t^2)^3 (1 - t^3)^4} \, .
\end{equation}
We see that $\dim K[V]^G = 7$ (corresponding to $5+5-3$ degrees of freedom). Thus there is one less physical parameter here as compared to the full (CP-odd) SM. This parameter would correspond to $\delta$ in the CKM matrix. Furthermore, the order-$4$ primary invariant was demoted to a secondary invariant.
The plethystic logarithm is given by
\begin{equation}
    \mathrm{PL} \left[ H(K[V]^G; u, d) \right] =
    (u^2 + u d + d^2) + (u^3 + d^3 + u^2 d + u d^2) + (u^2 d^2) - (u^6 d^6) \, .
\end{equation}
We see that also here there exists a syzygy of order $u^6d^6$, and we already know how it arises. 
The fact that $\left( J_{33} \right)^2 = 0$ suggests a relation between all (primary and secondary) invariants of the CP-even ring. 
We have explicitly confirmed that this relation is the syzygy in the CP-even ring. 

There is another computation we can perform. Let us divide both Hilbert series we have computed. This yields
\begin{equation}\label{eq:HS_CPV}
\frac{H_1}{H_2}\equiv
    \frac{H(K[V]^{\mathrm{SU}(3)}; u, d)}{H(K[V]^{\mathrm{SU}(2)}; u, d)} = 
    \frac{1}{1 - u^3 d^3} \, .
\end{equation}
This points to the source of CP violation in the quark sector, the parameter which is proportional to the Jarlskog invariant.
Finally, because  
\begin{equation}
    \mathrm{PL} \left[ H_1 / H_2 \right] = \mathrm{PL} \left[ H_1 \right] -
    \mathrm{PL} \left[ H_2 \right] \, ,
\end{equation}
it is trivial to compute the plethystic logarithm of eq.~\eqref{eq:HS_CPV} as
\begin{equation}
    \mathrm{PL} \left[\frac{H(K[V]^{\mathrm{SU}(3)}; u, d)}{H(K[V]^{\mathrm{SU}(2)}; u, d)} \right] = 
    u^3 d^3 \, ,
\end{equation}
thus explicitly exposing the CP-odd invariants of the original theory.
We stress that there seems to be no proof that the division of Hilbert series is always a Hilbert series in itself.
However, we certainly can always take the difference of plethystic logarithms of rings and their subrings in order to find
elements contained in one but not in the other.

\section{Birdtrack identities}\label{sec:birdtracks}
We mostly use the conventions of~\cite{Keppeler:2017kwt} with the following identities
\begin{align}
 &\birdtrack{16ex}{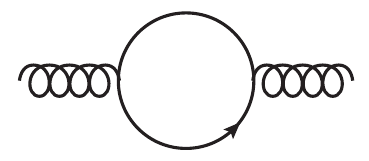}=T_{\rep{r}}\birdtrack{12ex}{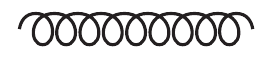}& &\text{with}& T_{\rep{r}}\delta^{ab} & =\Tr[t^at^b] \;,& \\
 &\birdtrack{16ex}{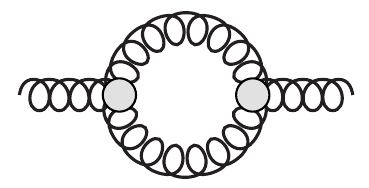}=C_D \birdtrack{12ex}{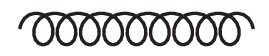}& &\text{with}& C_D & =\frac{N^2-4}{N}\;,& \\
 &\birdtrack{16ex}{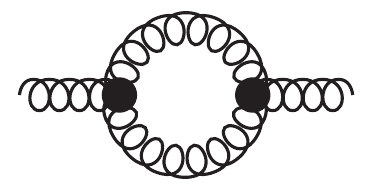}=C_A \birdtrack{12ex}{Diagrams/CA_short.pdf}& &\text{with}& C_A & =2T_{\rep{r}}N\;.& \\ 
 &\raisebox{6pt}{\birdtrack{16ex}{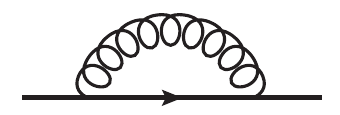}}=C_F \birdtrack{12ex}{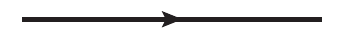}& &\text{with}& C_F & =T_{\rep{r}}\frac{N^2-1}{N}\;,&
\end{align}
\begin{equation}
 \birdtrack{16ex}{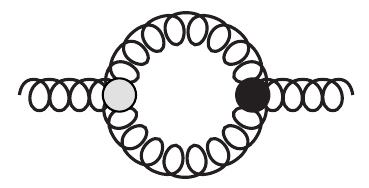}=\birdtrack{16ex}{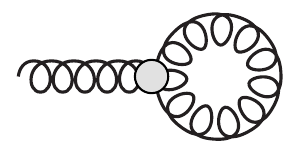}=\birdtrack{16ex}{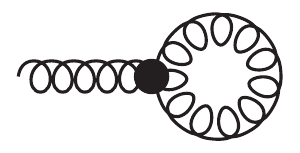}=0
\end{equation}

\section{Normalization of the projection operators}\label{sec:normalization}
The normalization of the orthogonal projection operators stated in section~\ref{sec:projector_basis}
is fixed by demanding idempotency as (automatically factorized and orthonormal) trivial singlet projection 
operators of \mbox{$\rep{8}^{\otimes k}\rightarrow\rep{8}^{\otimes k}$}.

To construct the corresponding operators, each of the operators of $\rep{8}^{\otimes k}\rightarrow\mathbbm{C}$ with $k$ legs is understood as \textit{half} a factorized projection operator in the mapping \mbox{$\rep{8}^{\otimes k}\rightarrow\rep{8}^{\otimes k}$}. To formally construct the full hermitian \mbox{$\rep{8}^{\otimes k}\rightarrow\rep{8}^{\otimes k}$} operator, 
all legs of the original operator are assigned as outgoing (effectively rotating the operator by $90^\circ$, flipping legs when necessary) and the hermitian conjugate of the operator (mirror image of operator with all fermion lines reverted) is included with all legs understood as ingoing (potential signs that originate from sometimes necessary permutation of some of the legs are irrelevant and absorbed in the normalization upon demanding idempotency). The thereby obtained operator is understood as a factorized $\rep{8}^{\otimes k}\rightarrow\rep{8}^{\otimes k}$ projection operator. The normalization is then fixed by demanding idempotency. For example, for the first diagram on the l.h.s.\ of eq.~\eqref{eq:8sa} the correct normalization $\mathcal{N}({\birdtrack{6ex}{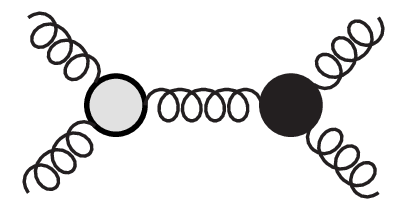}})$ is obtained by demanding
\begin{equation}
\mathcal{N}^2({\birdtrack{6ex}{Diagrams/8sa.pdf}})\;\birdtrack{14ex}{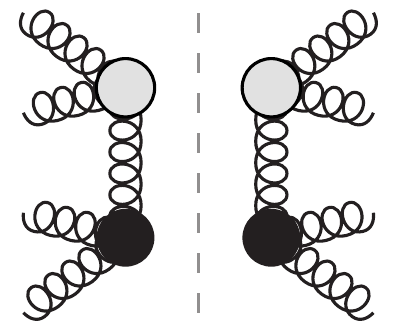}~\stackrel{!}{=}~
\mathcal{N}^4({\birdtrack{6ex}{Diagrams/8sa.pdf}})\;\birdtrack{24ex}{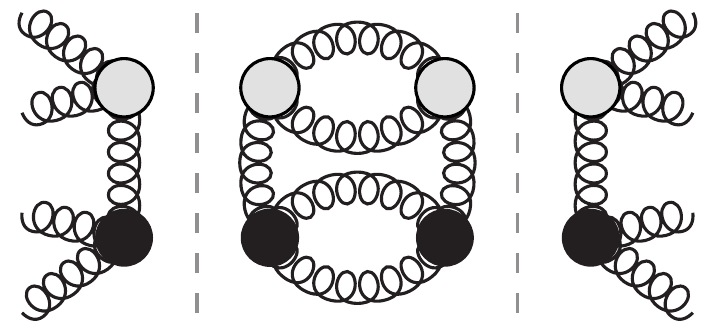}\;. 
\end{equation}
This fixes the norm to $\mathcal{N}({\birdtrack{6ex}{Diagrams/8sa.pdf}})=[C_D C_A (N^2-1)]^{-1/2}$. The trace of the projection operator on the l.h.s.\ is thereby 
automatically fixed to unity, confirming its property to project onto a one-dimensional subspace.
Analogous computations for all of the above projectors yield the normalization factors\footnote{%
For the $\rep{10}$, $\overline{\rep{10}}$, and $\rep{27}$ projection operators we only state the result specific for $N=3$ for simplicity.
More general expressions can be obtained, also for the last (in the case of $N=3$ vanishing) projection operator in $A\otimes A\rightarrow A\otimes A$, see e.g.~\cite{Keppeler:2012ih},
but we do not need these here.}
\begin{gather}
\mathcal{N}({\birdtrack{6ex}{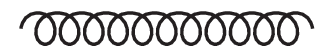}})=(N^2-1)^{-1/2}\;, \\
\mathcal{N}(\!\birdtrack{4ex}{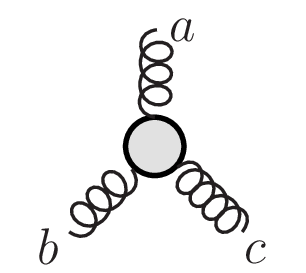}\!)=[C_D(N^2-1)]^{-1/2}\;,\qquad\mathcal{N}(\!\birdtrack{4ex}{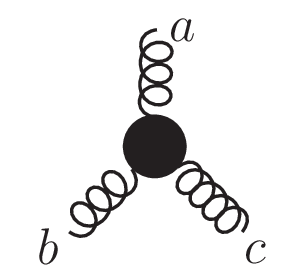}\!)=[2T_{\rep{r}}N(N^2-1)]^{-1/2}\;, \\
\mathcal{N}({\birdtrack{6ex}{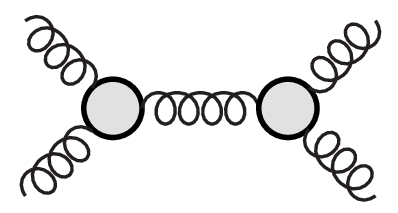}})=C_D(N^2-1)^{-1/2}\;, \qquad \mathcal{N}({\birdtrack{6ex}{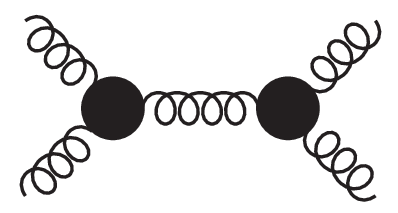}})=C_A(N^2-1)^{-1/2}\;, \\
\mathcal{N}({\birdtrack{6ex}{Diagrams/8sa.pdf}})=\mathcal{N}({\birdtrack{6ex}{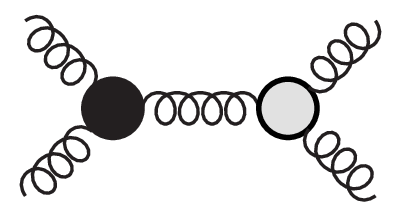}})=[C_D C_A (N^2-1)]^{-1/2}\;, \\
\mathcal{N}({\birdtrack{6ex}{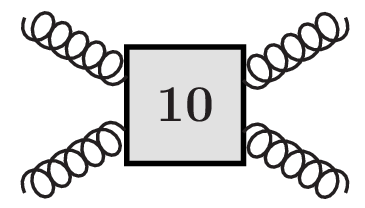}})=\mathcal{N}({\birdtrack{6ex}{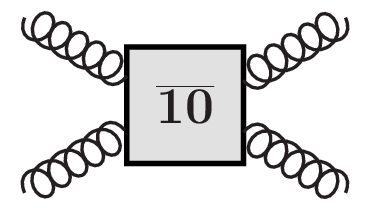}})=(10)^{-1/2} \;, \\
\mathcal{N}({\birdtrack{6ex}{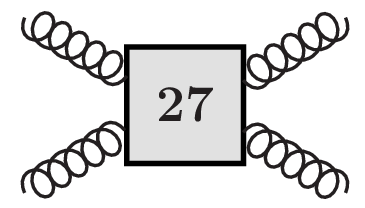}})=(27)^{-1/2} \;, \\
\mathcal{N}({\birdtrack{6ex}{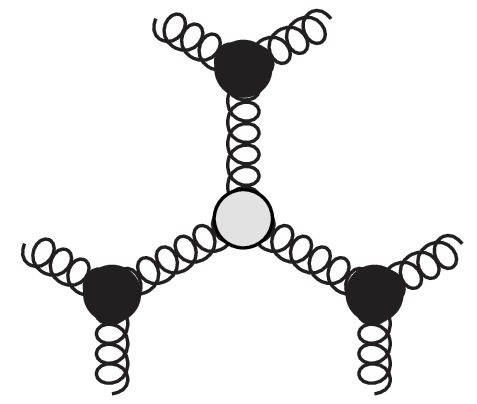}})= [C_A^3 C_D (N^2-1)]^{-1/2}\;.
\end{gather}

\section{Trace-based normalization of invariants and triangle plot}\label{App:Ialt}
\begin{figure}[!h!]
\centerline{\includegraphics[width=0.95\textwidth]{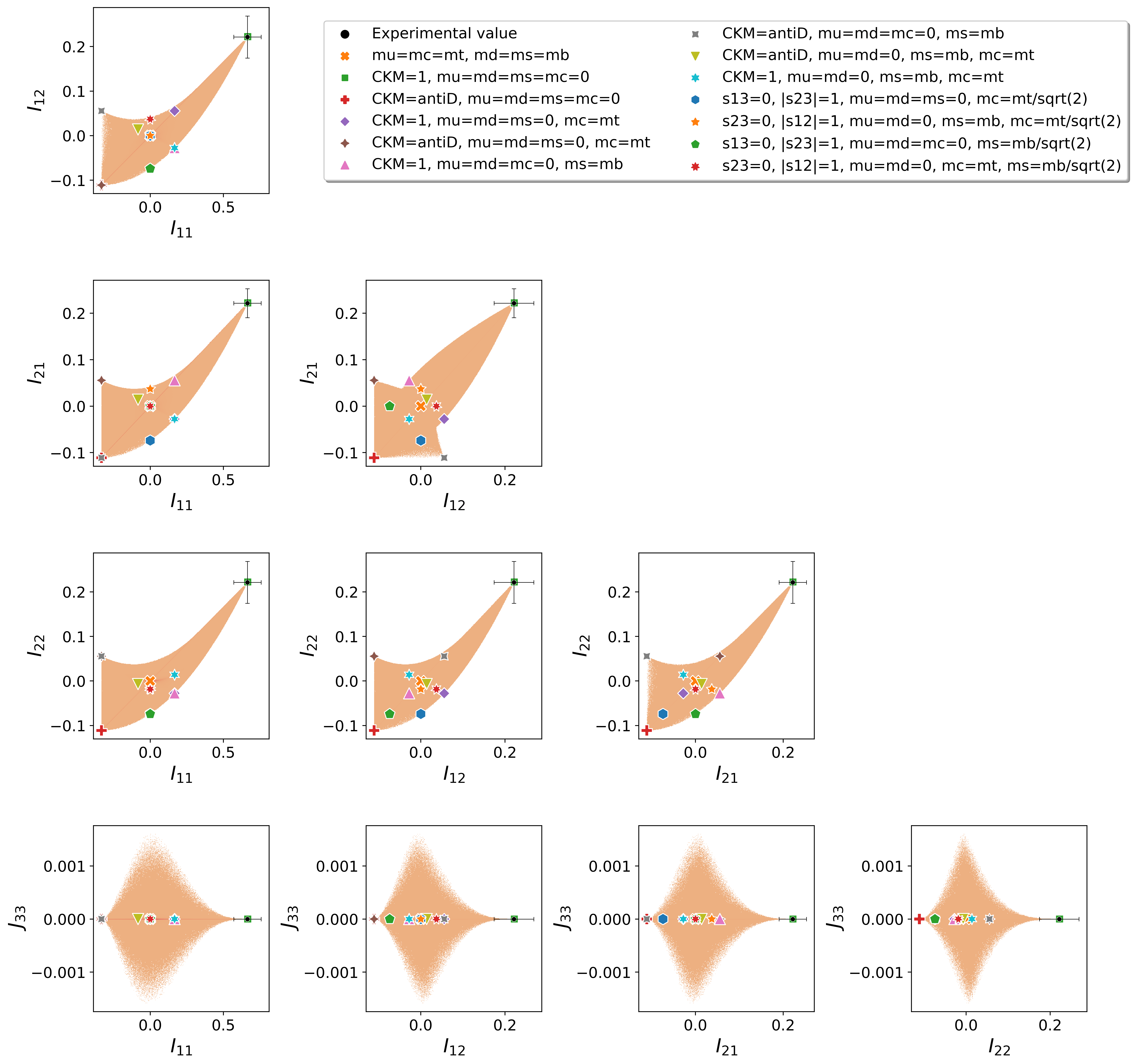}}
  \caption{\label{fig:triangleALT}
  Triangle correlations of the invariants $\hat{I}^{\mathrm{alt}}_{ij}$ normalized according to~\eqref{eq:ALTnormalization}.
  Otherwise the same as caption of figs.~\ref{fig:triangle} and~\ref{fig:highlight}.
  }
\end{figure}

As compared to eq.~\eqref{eq:normalization}, an arguably even ``more basis invariant'' choice is to normalize the invariants to the corresponding power of the up and down sector traces,
\begin{equation}\label{eq:ALTnormalization}
 \hat{I}^{\mathrm{alt}}_{ij}:=\frac{I_{ij}}{I_{10}^{i}\,I_{01}^{j}}\;.
\end{equation} 
For the experimentally determined location of the SM, the two different normalizations yield almost exactly the same numerical result for all invariants as on the r.h.s.\ of tab.~\ref{tab:invariants}. However, in less hierarchical regions of parameters, the analogue of figures~\ref{fig:triangle} and \ref{fig:highlight} for $\hat{I}^{\mathrm{alt}}_{ij}$ do, in fact, look very different and much less symmetrical, see the triangle correlation plot in figure~\ref{fig:triangleALT}. The measured location of the SM is even more pronounced as special. However, normalizing by the traces is less practical for scanning in degenerate regions of parameter space. Also, the apparent symmetric positioning of the special points seen in figures~\ref{fig:triangle} and~\ref{fig:highlight} is mostly lost.
\enlargethispage{1.5cm}

\section{Constraints on the parameter space of adjoint space invariants}\label{app:inner_product}
In this appendix we show that most of the invariants we use can be regarded as Frobenius inner products
of matrices. This allows us to derive bounds on the possible values of invariants and explore corners 
of the parameter space. Our results also confirms that the SM is a theory where $H_u$ and $H_d$, to first approximation, 
are linearly dependent. This is true because of two independent features: (i) The very large hierarchy between quark masses, and 
(ii) a CKM matrix that is close to the identity matrix. As a consequence, the SM is located near the corner of the parameter space
where all invariants take their maximal values.

\subsection{Frobenius inner product}
The Frobenius inner product $\expval{\cdot,\cdot}_F : V \times V \rightarrow \mathbbm{C}$ is defined as
\begin{equation}\label{eq:frobenius_inner}
    \expval{A,B}_F := \Tr \left( A^\dagger B \right) \, ,
\end{equation}
with the induced norm
\begin{equation}
    ||A||_F = \sqrt{\expval{A,A}_F} = \sqrt{\Tr \left( A^\dagger A \right)} \, .
\end{equation}
If the elements of $V$ are hermitian matrices then,
\begin{equation}
    \expval{A,B}_F = \Tr \left( A B \right) \, , \quad
    ||A||_F^2 = \Tr \left( A^2 \right) \, .
\end{equation}
Because \eqref{eq:frobenius_inner} is indeed an inner product, the usual
properties apply. In particular, the Cauchy-Schwarz inequality,
\begin{equation}
    \left|\expval{A,B}_F \right| \leq ||A||_F ||B||_F \, ,
\end{equation}
which in terms of traces of hermitian matrices is given by
\begin{equation}\label{eq:cauchy_tr}
    \left|\Tr \left( A B \right) \right| \leq \sqrt{\Tr \left( A^2 \right)}
    \sqrt{\Tr \left( B^2 \right)} \,.
\end{equation}
The inequality is saturated if and only if $A$ and $B$ are linearly dependent. 

\subsection{Adjoint space invariants as inner products and constraints}
Using the previous subsection, it is straightforward to identify our flavor invariants of eq.~\eqref{eq:trace_basis_inv} as Frobenius inner products.
We will use this here to derive bounds on their possible parameter space. 

A special feature of the traceless hermitian matrices is that
\begin{align}\label{eq:euclidean_mass}
    &I_{20} \equiv \Tr \left( H_u^2 \right) = \frac{1}{3} 
    \left[ (y_t^2 - y_c^2)^2 + (y_t^2 - y_u^2)^2 + (y_c^2 - y_u^2)^2 \right] \, , \nonumber \\
    &I_{02} \equiv \Tr \left( H_d^2 \right) = \frac{1}{3} 
    \left[ (y_b^2 - y_s^2)^2 + (y_b^2 - y_d^2)^2 + (y_s^2 - y_d^2)^2 \right] \, ,
\end{align}
where we recall that $y_{u,c,t}^2$ and $y_{d,s,b}^2$ are the strictly positive eigenvalues of the traceful matrices $\widetilde{H}_u$ and $\widetilde{H}_d$.
We see that $I_{20}$ and $I_{02}$ are an effective measure of the hierarchy in the quark Yukawa couplings. Normalizing the 
traces to their respective largest eigenvalue $y_t^2$ or $y_b^2$ (without loss of generality) as in eq.~\eqref{eq:normalization}, we get
\begin{align}\label{eq:euclidean_mass_norm}
    &\hat{I}_{20} \equiv \frac{\Tr \left( H_u^2 \right)}{y_t^4} = \frac{1}{3} 
    \left[ (1 - r_c)^2 + (1 - r_u)^2 + (r_c - r_u)^2 \right] \, , \nonumber \\
    &\hat{I}_{02} \equiv \frac{\Tr \left( H_d^2 \right)}{y_b^4} = \frac{1}{3} 
    \left[ (1 - r_s)^2 + (1 - r_d)^2 + (r_s - r_d)^2 \right] \, ,
\end{align}
with $r_{u,c} := (y_{u,c} / y_t )^2$ and $r_{d,s} := (y_{d,s} / y_b )^2$. It is then straightforward to check that
\begin{align}\label{eq:euclidean_mass_norm_max}
    &0 ~\leq~ \hat{I}_{20} ~\leq~\frac{2}{3}\,,& &\quad\text{and}& &0 ~\leq~ \hat{I}_{02} ~\leq~ \frac{2}{3}\,.&
\end{align}
Minimal and maximal values here corresponding to hierarchical Yukawas as
\begin{align}\label{eq:euclidean_mass_hier}
    &\max_{0\leq r_{u,d} \leq r_{c,s} \leq 1} 
    \hat{I}_{20},\;\hat{I}_{20} = \frac{2}{3}
    \quad \Rightarrow \quad r_{u,d}=0 \land \left( r_{c,s} = 0 
    \lor r_{c,s} = 1  \right) \, , \nonumber \\
    &\min_{0\leq r_{u,d} \leq r_{c,s} \leq 1} 
    \hat{I}_{20},\;\hat{I}_{02} = 0
    \quad \Rightarrow \quad  r_{u,d}=1 \land r_{c,s}=1 \, .
\end{align}
Thus, $\hat{I}_{20}$ and $\hat{I}_{02}$ are maximal when there is maximal hierarchy
and is minimal when there is degeneracy of all masses.

We can use eq.~\eqref{eq:cauchy_tr} to constrain (recall that all of our primary invariants obey $I_{ij} \in \mathbbm{R}$).
\begin{equation}\label{eq:cauchy_inv}
    |\hat{I}_{11}|  \leq \sqrt{\hat{I}_{20}}\; \sqrt{\hat{I}_{02}} \leq \frac{2}{3} \,.
\end{equation}
The Cauchy-Schwarz inequality is saturated if and only if $H_u$ and $H_d$ are linearly dependent,
\begin{equation}
    H_d = \lambda H_u \, , \quad \lambda \in \mathbbm{C} \, .
\end{equation}
Hence, sufficient condition for maximality of $I_{11}$ are 
\begin{equation}\label{eq:cauchy_i11_equa}
    y_{u,c}= 0 \, \land  y_{d,s} = 0 \,
    \land \, s_{13} = s_{23} = 0 \, ,
\end{equation}
The fact that the SM is to a good approximation fulfilling these conditions corresponds to the close-to maximality of the experimentally determined invariants.

For the pure cubic invariants one can derive
\begin{align}
    &\hat{I}_{30} \equiv \frac{\Tr \left( H_u^3 \right)}{y_t^6} = \frac{1}{9} 
    \left[ (2 - r_c^2-r_u^2)(1 + r_u^2 - 2 r_c^2 )(1 + r_c^2 - 2 r_u^2) \right] \, , \nonumber \\
    &\hat{I}_{03} \equiv \frac{\Tr \left( H_d^3 \right)}{y_b^6} = \frac{1}{9} 
    \left[ (2 - r_s^2-r_d^2)(1 + r_d^2 - 2 r_s^2 )(1 + r_s^2 - 2 r_d^2) \right] \, ,
\end{align}
which straightforwardly allows to show 
\begin{align}
    &-\frac29 ~\leq~ \hat{I}_{30} ~\leq~\frac29\,,& &\quad\text{and}& &-\frac29 ~\leq~ \hat{I}_{03} ~\leq~ \frac29\,.&
\end{align}

An exact bound can be derived for 
\begin{equation}
\left|\frac{\Tr(H_u^2H_d^2)}{y_t^4\,y_b^4}\right|  \leq \frac12 \sqrt{\hat{I}_{20}}\; \sqrt{\hat{I}_{02}} \leq \frac{2}{9} \,,
\end{equation}
upon noting that $\mathrm{Tr}(H_{u,d}^{4}) = \frac{1}{2} \mathrm{Tr}( H_{u,d}^{2} )^2$ by the Cayley-Hamilton theorem. Using these results, also
\begin{equation}
 \hat{I}_{22} \leq \frac29\;.
\end{equation}
For the remaining non-trivial invariants, $\hat{I}_{21}\leq\frac29$ and $\hat{I}_{12}\leq\frac29$, exact bounds can be derived from careful treatment of the $d$ tensor inner product, but for now we refer to the numerical proof of their boundedness shown in figs.~\ref{fig:triangle} and~\ref{fig:triangleALT}.

\section{Running of CKM parameters}\label{app:CKMrunning}
From the running of $\widetilde{H}_u$ and $\widetilde{H}_d$, it is possible to extract the running of $|V_{ub}|$, $|V_{cb}|$, $|V_{td}|$ and $J$.
This appendix reproduces the running of CKM parameters derived in ref.~\cite{Babu:1987im} (see also \cite{Sasaki:1986jv, Grossman:2022ehc} and references therein) but using updated values for the low energy CKM parameters. We show the results here both for completeness, and as an important cross-check to confirm the correctness of the running of our invariants.

\begin{figure}[t]
\begin{minipage}[c]{0.5\textwidth}
\centerline{\includegraphics[width=1.0\textwidth]{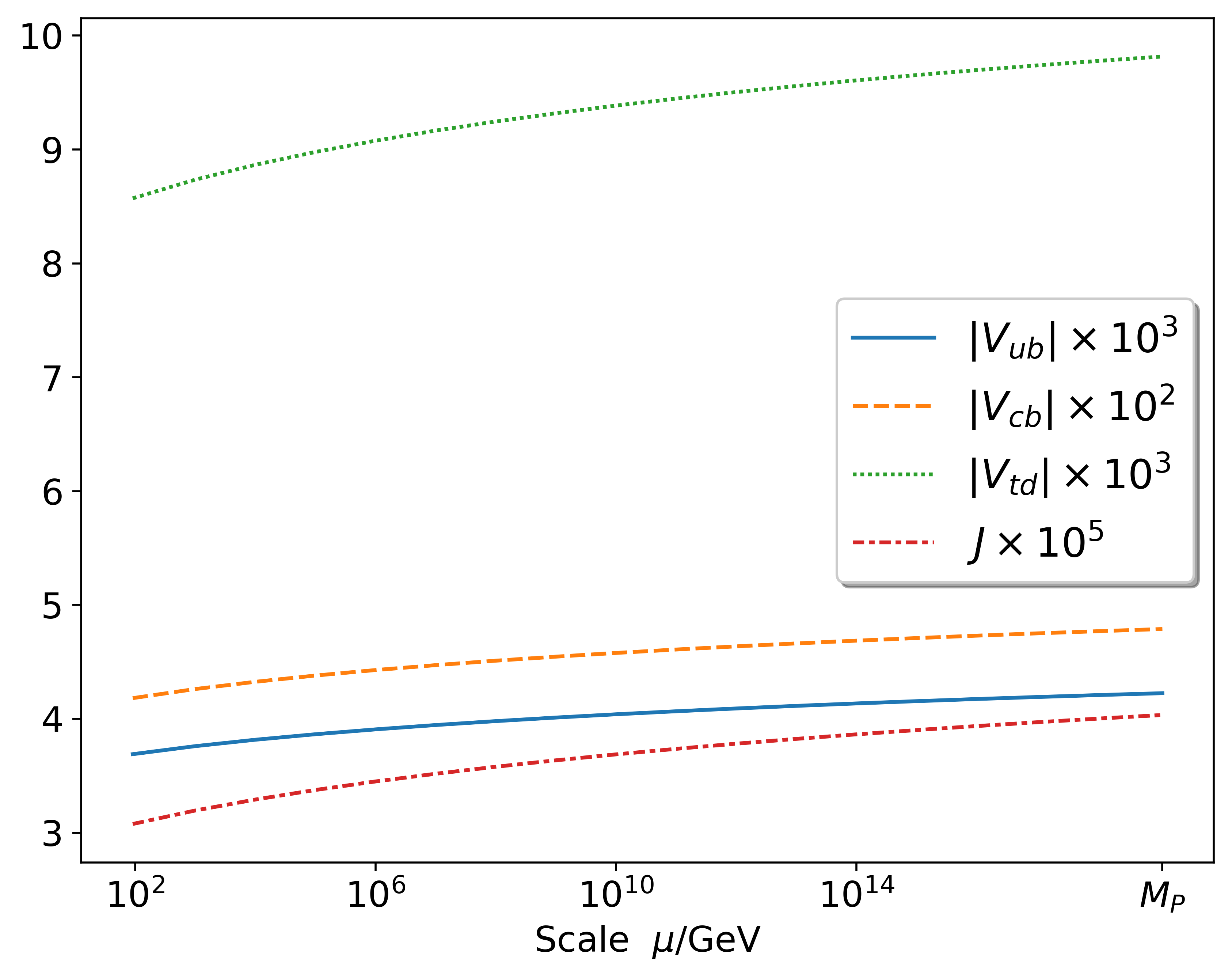}}
\end{minipage}%
\begin{minipage}[c]{0.5\textwidth}
\centerline{\includegraphics[width=1.0\textwidth]{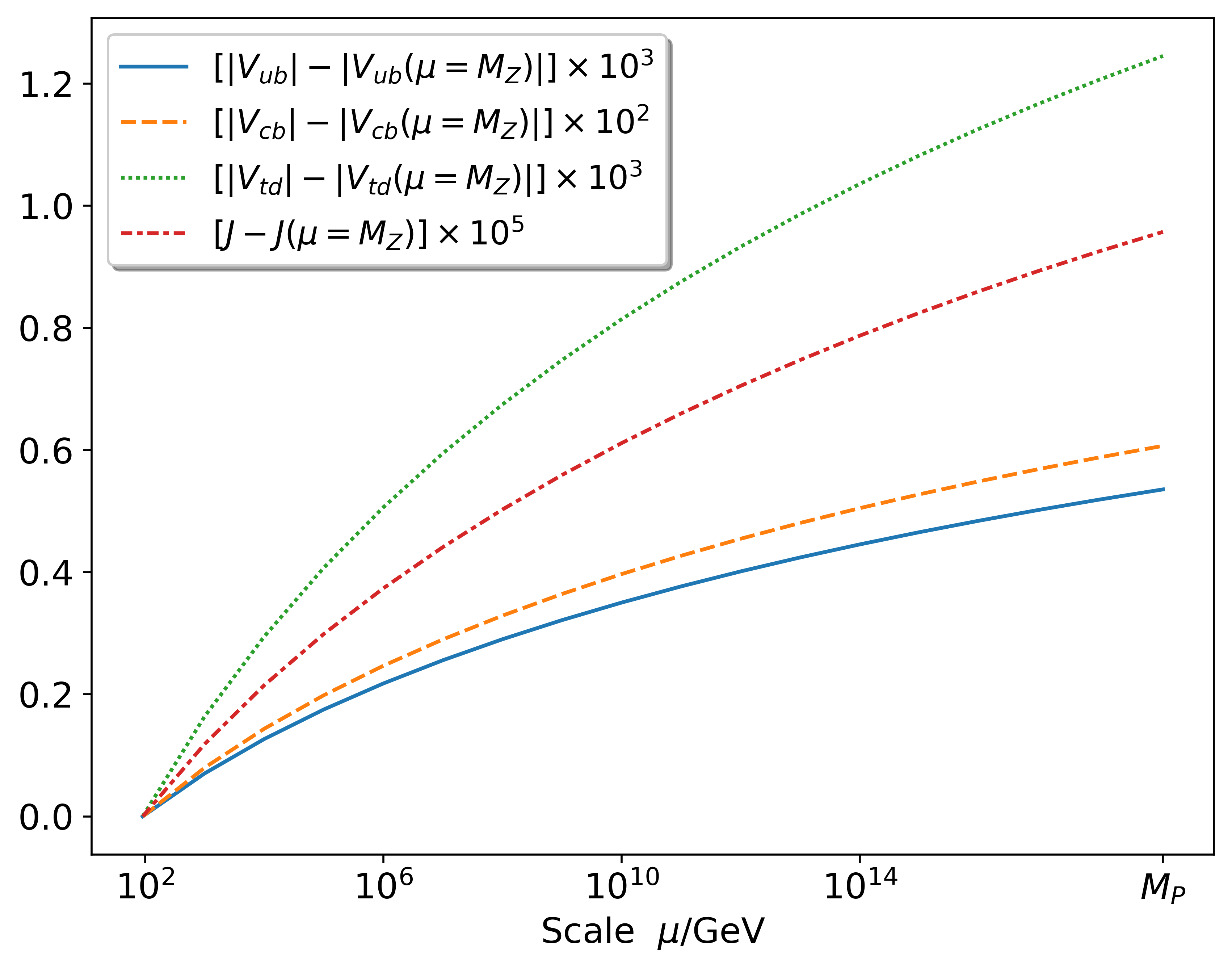}}
\end{minipage}
\caption{\label{fig:ckm_running}
Running of the CKM elements $|V_{ub}|$, $|V_{cb}|$, $|V_{td}|$, and running of $J$ (left) as a function of the renormalization scale $\mu$. 
On the r.h.s.\ we show the change of their values as compared to the electroweak scale $\mu = M_Z$.}
\end{figure}

At the scale $\mu = M_Z$ we begin the running by choosing a basis
where $H_u$ is diagonal.\footnote{Since $\widetilde{H}_{u,d}$ and $H_{u,d}$ are simultaneously diagonalized one can use both, $\widetilde{H}_{u,d}$ or $H_{u,d}$, in order to extract the running of the CKM parameters.
Without loss of generality we use $H_{u,d}$ in this appendix.}
At any scale, the CKM matrix is defined as the matrix that diagonalizes $H_d$ in the basis where $H_u$ is diagonal (see eq.~\eqref{eq:HuPhysical}). 
As we run to higher values of the scale $\mu$, $H_u(\mu)$ evolves to a different, in general, non-diagonal basis. This contains the running of the physical parameters, in addition to disguising them by an inconvenient basis choice (this is one of the reason why running directly in the invariants is superior -- unphysical effects like rotations of the basis drop out by construction). At any given scale, we can extract the equations
\begin{align}
    H_u(\mu) = V_{u,\mathrm{L}}(\mu)\, D_u^2(\mu)\, V_{u,\mathrm{L}}^\dagger(\mu) \, , \nonumber \\
    H_d(\mu) = V_{d,\mathrm{L}}(\mu)\, D_d^2(\mu)\, V_{d,\mathrm{L}}^\dagger(\mu) \, ,
\end{align}
where $D_{u,d}^2 = \mathrm{diag} (y_{u,d}^2, y_{c,s}^2, y_{t,b}^2)$. The corresponding CKM matrix at that scale $\mu$ is then given as 
\begin{equation}
    V_{\mathrm{CKM}}(\mu) = V_{u,\mathrm{L}}^\dagger(\mu) \, V_{d,\mathrm{L}}(\mu) \, .
\end{equation}
This allows us to extract the values of $|V_{\mathrm{CKM}}(\mu)|$ and $J(\mu)$, here using the definition $J= \Im \left( V_{ud} V_{cs} V_{us}^* V_{cd}^* \right)$.
We show their evolution in figure~\ref{fig:ckm_running}, which should be compared to figure~1~of~\cite{Babu:1987im}. We also confirm the running of the CKM parameter $A$ as reported in~\cite{Grossman:2022ehc}. Explicitly we find $A(10^{15}\,\mathrm{GeV})\approx0.930$, $A(10^{19}\,\mathrm{GeV})\approx0.945$
and virtually no running of $\lambda$, $\eta$, $\rho$ which only change at the relative order of $10^{-4}$.

\bibliographystyle{JHEP}
\enlargethispage{1cm}
\bibliography{bibliography}

\end{document}